\documentclass[twocolumn]{aastex61}
\pdfoutput=1 
\usepackage{amsmath,amstext}
\usepackage[T1]{fontenc}
\usepackage{apjfonts} 
\usepackage[figure,figure*]{hypcap}
\usepackage{subfigure}
\usepackage{longtable,booktabs}
\usepackage{threeparttablex}
\usepackage{multirow}
\usepackage{natbib} 
 
\def\gtsima{$\; \buildrel > \over \sim \;$}
\def\ltsima{$\; \buildrel < \over \sim \;$}
\def\gsim{\lower.5ex\hbox{\gtsima}}
\def\lsim{\lower.5ex\hbox{\ltsima}}
\def\simgt{\lower.5ex\hbox{\gtsima}}
\def\simlt{\lower.5ex\hbox{\ltsima}}
\def\simpr{\lower.5ex\hbox{\prosima}}

\shorttitle{EoR LF faint-end constraints from FFs}
\shortauthors{Yue et al.}

\begin{document}

\title{On the faint-end of the galaxy luminosity function in the Epoch of Reionization: updated constraints from the {\it HST} Frontier Fields}

\author{B. Yue}
\affiliation{National Astronomical Observatories, Chinese Academy of Sciences, 20A Datun Road, Chaoyang District, Beijing, 100012, China}
\author{M. Castellano}
\affiliation{INAF - Osservatorio Astronomico di Roma, Via Frascati 33, I -00078Monte Porzio Catone (RM), Italy}
\author{A. Ferrara}
\affiliation{Scuola Normale Superiore, Piazza dei Cavalieri 7, I-56126 Pisa, Italy}
\affiliation{Kavli IPMU (WPI), Todai Institutes for Advanced Study, the University of Tokyo, Japan}
\author{A. Fontana}
\affiliation{INAF - Osservatorio Astronomico di Roma, Via Frascati 33, I -00078Monte Porzio Catone (RM), Italy}
\author{E. Merlin}
\affiliation{INAF - Osservatorio Astronomico di Roma, Via Frascati 33, I -00078Monte Porzio Catone (RM), Italy}
\author{R. ~Amor\'{\i}n}
\affiliation{Cavendish Laboratory, University of Cambridge, 19 JJ Thomson Avenue, Cambridge, CB3 0HE, UK}
\affiliation{Kavli Institute for Cosmology, University of Cambridge, Madingley Road, Cambridge CB3 0HA, UK}
\author{A. Grazian}
\affiliation{INAF - Osservatorio Astronomico di Roma, Via Frascati 33, I -00078Monte Porzio Catone (RM), Italy}
\author{E. ~M\'armol-Queralto}
\affiliation{SUPA, Scottish Universities Physics Alliance, Institute for Astronomy, University of Edinburgh, Royal Observatory, Edinburgh, EH9 3HJ, U.K.}
\author{M.~J.~Micha{\l}owski}
\affiliation{Astronomical Observatory Institute, Faculty of Physics, Adam Mickiewicz University, ul.~S{\l}oneczna 36, 60-286 Pozna{\'n}, Poland}
\author{A. ~Mortlock}
\affiliation{SUPA, Scottish Universities Physics Alliance, Institute for Astronomy, University of Edinburgh, Royal Observatory, Edinburgh, EH9 3HJ, U.K.} 
\author{D. Paris}
\affiliation{INAF - Osservatorio Astronomico di Roma, Via Frascati 33, I -00078Monte Porzio Catone (RM), Italy}
\author{S. ~Parsa}
\affiliation{SUPA, Scottish Universities Physics Alliance, Institute for Astronomy, University of Edinburgh, Royal Observatory, Edinburgh, EH9 3HJ, U.K.} 
\author{S. Pilo}
\affiliation{INAF - Osservatorio Astronomico di Roma, Via Frascati 33, I -00078Monte Porzio Catone (RM), Italy}
\author{P. ~Santini}
\affiliation{INAF - Osservatorio Astronomico di Roma, Via Frascati 33, I -00078Monte Porzio Catone (RM), Italy}
\author{M. Di Criscienzo}
\affiliation{INAF - Osservatorio Astronomico di Roma, Via Frascati 33, I -00078Monte Porzio Catone (RM), Italy}

\begin{abstract}

Ultra-faint galaxies are hosted by small dark matter halos with shallow gravitational potential wells, hence their star formation activity is more sensitive to feedback effects. The shape of the faint-end of the high-$z$ galaxy luminosity function (LF) contains important information on star formation  and its interaction with the reionization process during the Epoch of Reionization (EoR). High-$z$ galaxies with $M_{\rm UV}\gsim-17$ have only recently become accessible thanks to the Frontier Fields (FFs) survey combining deep {\it HST} imaging and the gravitational lensing effect. In this paper we investigate the faint-end of the LF at redshift $>$5 using the data of FFs clusters Abell 2744 (A2744), MACSJ0416.1-2403 (M0416), MACSJ0717.5+3745 (M0717) and MACSJ1149.5+2223 (M1149). We analyze both an empirical and a physically-motivated LF model to obtain constraints on a possible turn-over of LF at faint magnitudes. In the empirical model the LF drops fast when the absolute UV magnitude $M_{\rm UV}$ is much larger than a turn-over absolute UV magnitude $M_{\rm UV}^{\rm T}$. We obtain $M_{\rm UV}^{\rm T}\gsim-14.6 $ (15.2) at 1 (2) $\sigma$ confidence level (C.L.) for $z\sim6$. In the physically-motivated analytical model, star formation in halos with circular velocity below $v_c^*$ is fully quenched if these halos are located in ionized regions. Using updated lensing models and new additional FFs data, we re-analyze previous constraints on $v_c^*$ and $f_{\rm esc}$ presented by Castellano et al. 2016a (C16a) using a smaller dataset. 
We obtain new constraints on $v_c^*\lsim 59$ km s$^{-1}$ and $f_{\rm esc}\lsim 56\%$ (both at 2$\sigma$ C.L.) 
and conclude that there is no turn-over detected so far from the analyzed FFs data. Forthcoming {\it JWST} observations will be key to tight these constraints further.

\end{abstract}

\keywords{dark ages, reionization, first stars --- galaxies: high-redshift --- gravitational lensing: strong}

\section{Introduction}

During the Epoch of Reionization (EoR, $6\lsim z \lsim 30$), the intergalactic medium (IGM) was gradually ionized by energetic photons mainly emitted by the first galaxies. This in turn leads to the suppression of star formation in small galaxies, because their host halos hardly collect  gas from ionized environment. This feedback effect raises the following questions: How faint the first galaxies could be and which halos could sustain star formation activity during the EoR?

According to the hierarchical structure formation scenario, smaller dark matter halos are much more common than bigger ones in the Universe, resulting in an overwhelming numerical abundance of very faint galaxies \citep{Mason2015,Mashian2016,2017MNRAS.464.1633F,2016MNRAS.462..235L}. Thereby, faint galaxies are promising candidates as main sources of reionizing photons \citep[e.g.][]{2015ApJ...811..140B,Robertson2015ApJ,Castellano2016ApJ}, with a crucial contribution possibly coming from objects far below the detection limits of even the deepest existing surveys \citep{Salvaterra2011, Robertson2013,2008MNRAS.385L..58C, 2007MNRAS.380L...6C,2013MNRAS.434.1486D,2013MNRAS.429.2718S}. Moreover, faint galaxies are less clustered and their environment gas is less clumped, therefore they are more effective in reionizing the IGM.

To understand the role of star-forming galaxies in the reionization process, it is thus crucial to constrain their number density and star-formation efficiency by studying the UV luminosity function (LF) down to the faintest limits. The faint-end of the UV LF at high redshift has been found to have a steep slope at least down to absolute UV magnitudes $M_{\rm UV}\sim-16$ (\citealt{McLure2013,Bouwens2015}, B15 hereafter) in blank fields or even $M_{\rm UV}\sim-12$ in gravitational lensing fields (\citealt{Livermore2017}, however see \citealt{Bouwens2017-FFs}, these two are L17 and B17 hereafter). The high-$z$ LFs have also been reconstructed from the number of ultra-faint dwarf galaxies \--- some of which are believed fossils of reionization galaxies \--- in the Local Group. For example, \citet{2014ApJ...794L...3W} conclude that the LF at $z\sim5$ does not have any break at least down to $M_{\rm UV}\sim-10$. At the same time, the detection of a break or turn-over would have an important implication related to the nature of the first galaxies and their contribution to reionization \citep[e.g.][]{2015A&A...578A..83G,Madau2015,Mitra2016}.
 
Star formation in dark matter halos relies on the gas cooling process. However, both supernova explosion and ionizing radiation could prevent cooling. These feedback effects reduce the efficiency of star formation \citep{2013MNRAS.434.1486D,2016ApJ...833...84X,2016MNRAS.460..417S} or even completely quench it, if the halo mass is too small. As a result, the number of galaxies hosted by small halos drops and we expect to see a ``turn-over" in the faint-end of the galaxy LFs (\citealt{Yue2016-faint-end}, Y16 hereafter). For example, in the ``Cosmic Reionization on Computers" (CROC) project,  by galaxy formation and radiative transfer numerical simulations, it is found that the UV LF turns over at $M_{\rm UV}\sim-14$ to $\sim-12$ \citep{Gnedin2016-faint-end}. And in the ``FirstLight" project with radiative feedback effects, the LF has a flattening at $M_{\rm UV}\gsim-14$ (with host halos' circular velocity $\sim30$ - 40 km s$^{-1}$, \citealt{Ceverino2017_FirstLight}). Modifications of the initial power spectrum as in WDM cosmologies can also have a similar effect, see e.g. \citet{Dayal2015,Menci2016,Menci2017}. Observations of galaxies around the turn-over would greatly increase our knowledge of the star formation physics in galaxies contributing most to reionization, and may directly answer our questions in the first paragraph \citep{Yue2014-FFs}.

Until now, there is no evidence that confirms or rules out the existence of such turn-over in both regular surveys and in gravitational lensing surveys, probably because the turn-over magnitude is still fainter than the limiting magnitudes of current measurements, see e.g. \citet{McLure2013}; B15; \citet{Atek2015_18A}; \citet{Atek2015_69A} (A15 hereafter); L17; \citet{Laporte2016,Ishigaki2017} (I17 hereafter). 

With the help of strong magnification effects, the gravitational lensing provides an opportunity to detect galaxies below the detection limits of regular surveys. However, the cost is that the survey volume is reduced, and lensing models introduce extra uncertainties into the recovered intrinsic brightness of observed galaxies (B17).

The Frontier Fields (FFs) survey observed six massive galaxy clusters and their parallel fields in optical and near-infrared bands with the {\it Hubble}\footnote{\url{http://www.stsci.edu/hst/campaigns/frontier-fields/}}  and {\it Spitzer}\footnote{\url{http://ssc.spitzer.caltech.edu/warmmission/scheduling/approvedprograms/ddt/frontier/}} space telescopes \citep{Lotz2017-FFs}. These observations were also followed up by other observatories at longer and shorter wavelengthes, e.g. {\it ALMA} \citep{ALMA-FFs-I,ALMA-FFs-III} and {\it Chandra} \citep{Chandra-FFs-2015, Chandra-FFs-2016,Chandra-FFs-2017}. Using the clusters as lenses, these images are deep enough to unveil  faint galaxy populations at the EoR. 

In Y16, we have derived the form of the LF faint-end during and after the EoR by assuming that the star formation in halos with circular velocity below a threshold $v_c^*$ and located in ionized bubbles is quenched, where $v_c^*$ is a free parameter. In \citet{Castellano2016a} (C16a hereafter) we constrained $v_c^*\lsim 60$ km s$^{-1}$ ($2\sigma$ C.L.) by using the observed number counts of ultra-faint galaxies in two of the six FFs cluster fields, A2744 and M0416, and the {\it Planck}2015 results for $\tau$ \citep{2016A&A...594A..13P}.

Recently, by using two FFs clusters Abell 2744 (A2744) and MACSJ0416.1-2403 (M0416),  L17 found that the faint-end of the LF at $z\sim6$ always has steep slope ($\alpha\sim-2$) and does not turn over at $\gsim-12.5$. Generally it is expected that at higher redshift the turn over magnitude is fainter, because at the earlier reionization stage the 
radiative feedback effects should be weaker, and halos at higher redshift are more concentrated and easier to hold their gas.  
Therefore from L17 results it can be reasonably inferred that during the EoR ($z>6$), the LF faint-end slope is steep even at magnitudes fainter than $-12.5$. If this is the case, the galaxies that ionized the Universe have already been uncovered \citep{Robertson2013, Robertson2015ApJ}. 
However, their result was questioned by B17 who argued that L17 may overestimate the volume density at the faint end due to: 1) an excess of sources near the completeness limit; and 2) the assumption of too large intrinsic half-light radii.

B17 investigated the impact of magnification errors on the LF carefully and found that at $M_{\rm UV}\gsim -14$ the 
systematic differences of magnifications from different lensing models are extremely high.
They developed a new model that incorporates the magnification errors into the LF, and by analyzing four FFs clusters: A2744, M0416 plus MACSJ0717.5+3745 (M0717) and MACSJ1149.5+2223 (M1149) they obtained the constraints that the LF should not turn over at least at $M_{\rm UV}<-15.3$ to $-14.2$ (1$\sigma$ C.L.), consistent with C16a.

In this paper, we expand the analysis presented in C16a by adding new FFs data and improved lensing models to obtain number counts in the two additional FFs clusters and update the previous two clusters.  Throughout this paper we use the following cosmological parameters: $\Omega_m=0.308, \Omega_\Lambda=0.692, \Omega_b=0.048, h=0.678, \sigma_8=0.815, n_s=0.97$ \citep{2016A&A...594A..13P}, magnitudes are presented in AB system.

\section{Methods}

\subsection{Observations}

The photometric catalogues of high-$z$ galaxies used in the present paper are provided by the ASTRODEEP team (\citealt{Castellano2016c,Merlin2016b,DiCriscienzo2017}), and all the lensing models are provided by the FFs team on the project website\footnote{\url{https://archive.stsci.edu/prepds/frontier/lensmodels/}}.

The high-$z$ sample comprises all sources with $H_{\rm 160,int}\ge 27.5$ from the ASTRODEEP catalogs of FFs clusters A2744, M0416 \citep{Merlin2016b,Castellano2016a}, M0717 and M1149 \citep{DiCriscienzo2017}\footnote{Download: \url{http://www.astrodeep.eu/frontier-fields-download/};\\ Catalogue interface: \url{http://astrodeep.u-strasbg.fr/ff/index.html}}, where $H_{\rm 160,int}$ is the demagnified apparent magnitude at the {\it HST} F160W band (H band).
The model described in Sec. \ref{empirical} will use sample galaxies with $5.0<z<7.0$, while the model described in Sec. \ref{physical} will use sample galaxies with $5.0<z<9.5$, see details in the relevant sections. The original samples presented by ASTRODEEP team have redshifts up to $\sim10$. However for objects $z\gsim9.5$, their redshifts may be not correctly measured. Moreover, samples with $z\gsim9.5$ are only detectable in one band. Considering these reasons we do not select samples with $z>9.5$.

All catalogues include photometry from the available {\it HST} ACS and WFC3 bands \citep[B435, V606, I814, Y105, J125, JH140, H160, see e.g.][]{Lotz2017-FFs} and from deep K-band \citep{brammer2016} and IRAC 3.6 $\mu$m and 4.5 $\mu$m data (PI Capak).
Sources are detected on the H160 band after removing foreground light both from bright cluster galaxies and the diffuse intra-cluster light (ICL) as described in detail in \citet{Merlin2016b}. Foreground light is also removed from the {\it HST} bands before estimating photometry with \verb|SExtractor| \citep{Bertin1996} in dual-image mode. Photometry from the lower resolution Ks and IRAC images has been obtained with \verb|T-PHOT| v2.0 \citep{Merlin2015,Merlin2016a}.  
Photometric redshifts for all the sources have been measured with six different techniques based on different codes and assumptions. The FFs sources are then assigned the median of the six available photometric redshift estimates in order to minimize systematics and improve the accuracy. 
The final typical error on the photo-$z$ is $\sim0.04\times(1+z)$ \citep{Castellano2016a,DiCriscienzo2017}. 
For the four clusters A2744, M0416, M0717 and M1149, the apparent magnitudes brighter than which more than 10\% point-like objects could be successfully resolved are 28.8, 28.8, 28.5 and 28.7 respectively.

\begin{figure}
\centering{
\includegraphics[scale=0.4]{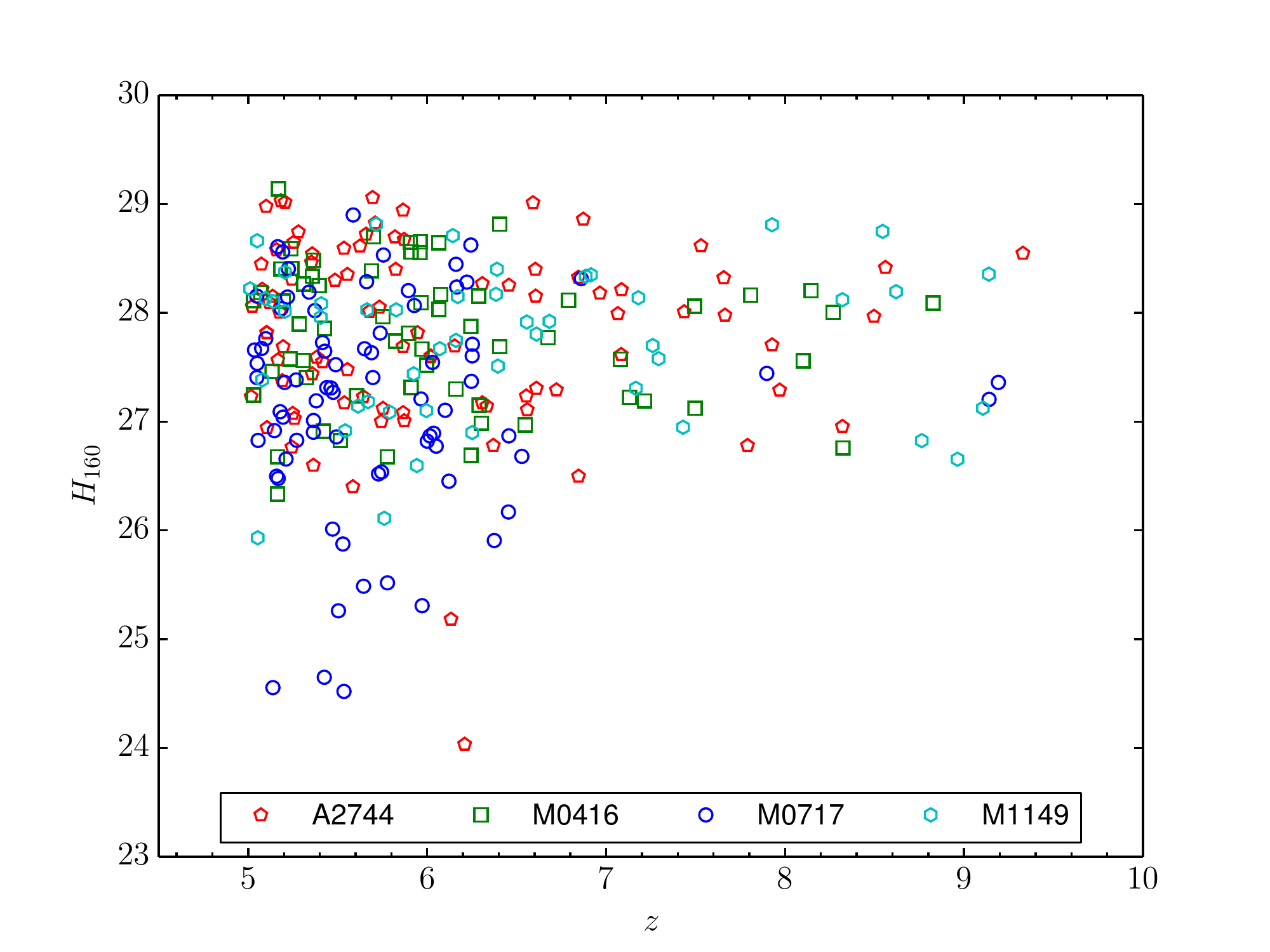}
\includegraphics[scale=0.4]{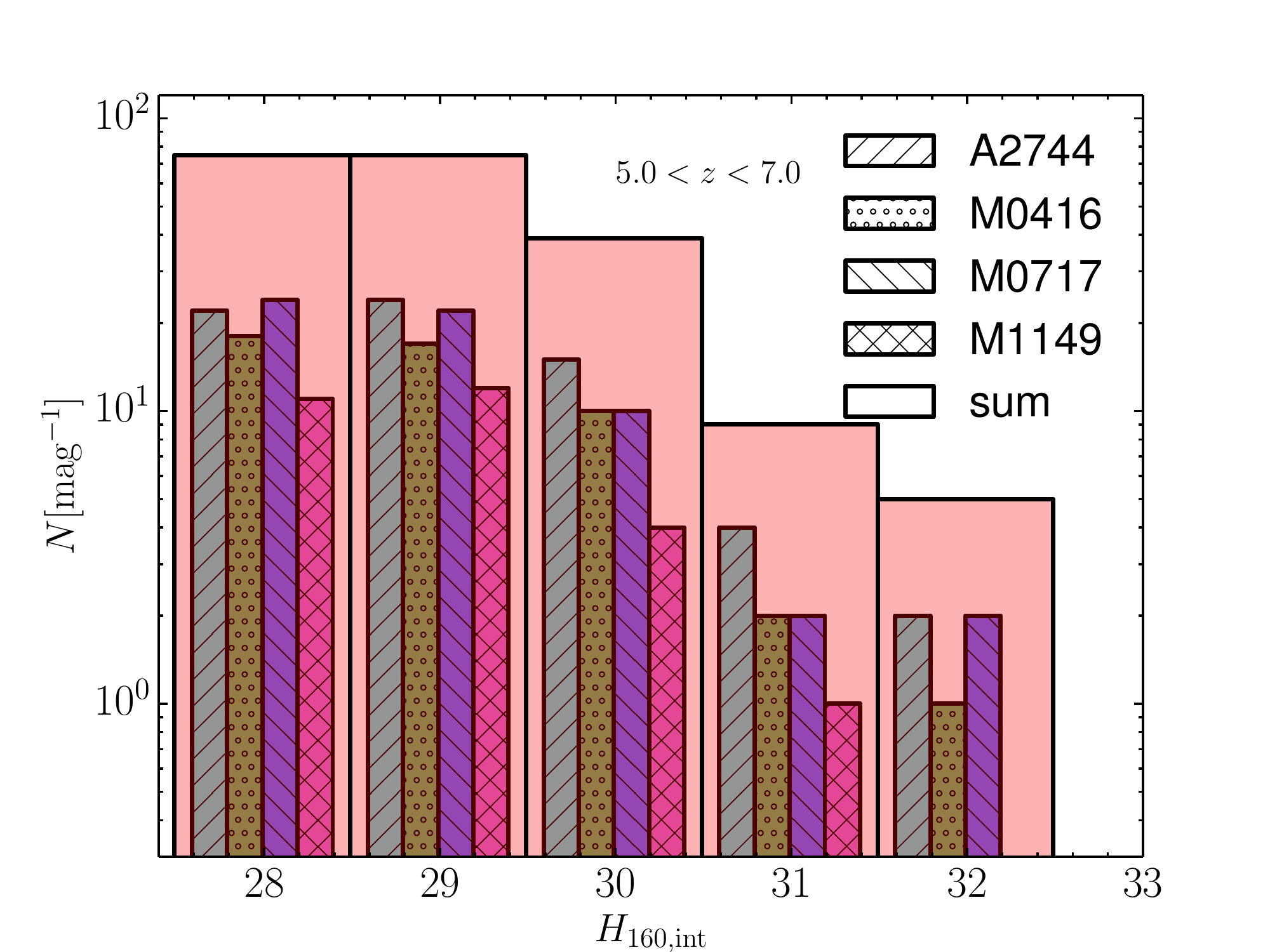}
\includegraphics[scale=0.4]{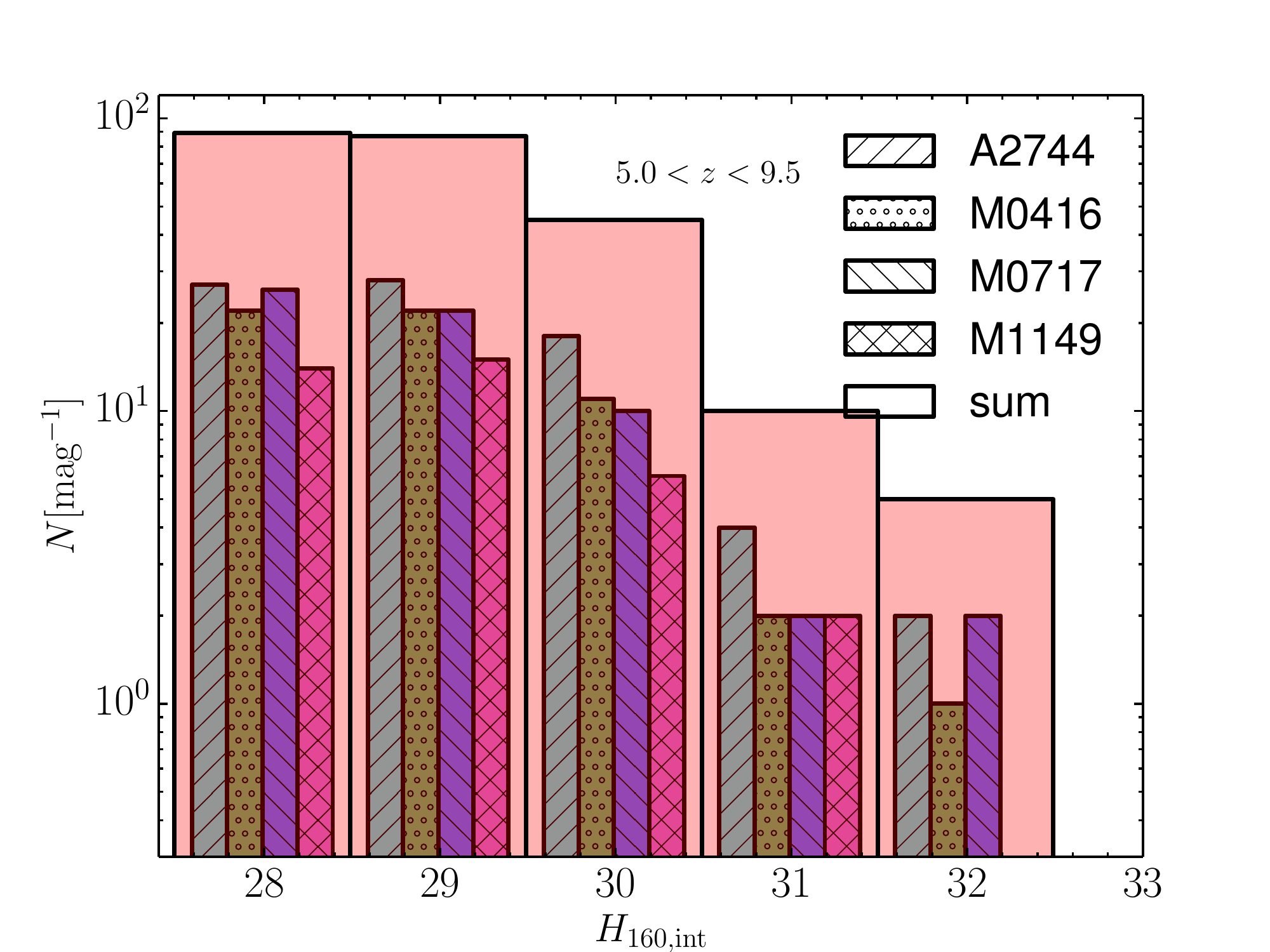}
\caption{
{\it Top:} The observed H band magnitudes vs. redshift of the sample galaxies with photometric redshifts between 5.0 and 9.5 in the four FFs clusters respectively.
{\it Middle and bottom:} The galaxy number counts between $z=5.0-7.0$ and $z=5.0-9.5$ vs. demagnified H band apparent magnitude for the four FFs clusters. For each cluster we plot the median of the number counts reconstructed using lensing models listed in Tab. \ref{tb:clusters}. We also plot the sum of the four clusters. Histograms in same group are in the same magnitude bin, for displaying purpose we shift their x-coordinates.
}
\label{fig:numbercount_obs}
}
\end{figure}

In the top panel of Fig. \ref{fig:numbercount_obs} we plot the observed H band apparent magnitude, $H_{160}$, vs. redshift for our selected sample galaxies 
(galaxies with photometric redshift between 5.0 and 9.5 and with demagnified H band magnitude larger than 27.5 in either of lensing models) in the four FFs clusters.  
There are 73 (87), 51 (62), 73 (76) and 34 (47) galaxies with $5.0<z<7.0$ ($5.0<z<9.5$) in clusters A2744, M0416, M0717 and M1149 respectively.
In the Tab. \ref{tb:gal_list} in the Appendix \ref{gal_list} we list the unique ID in the ASTRODEEP catalog of all these objects, so that their properties like the released SEDs could be found directly in the website.

The magnification for each observed source is estimated on the basis of the relevant photometric redshift from  shear and mass surface density values at its barycenter of the light distribution. All models made available on  the STSCI website \footnote{\url{http://www.stsci.edu/hst/campaigns/frontier-fields/Lensing-Models}} are used.

\begin{figure*}
\centering{
\includegraphics[scale=0.8]{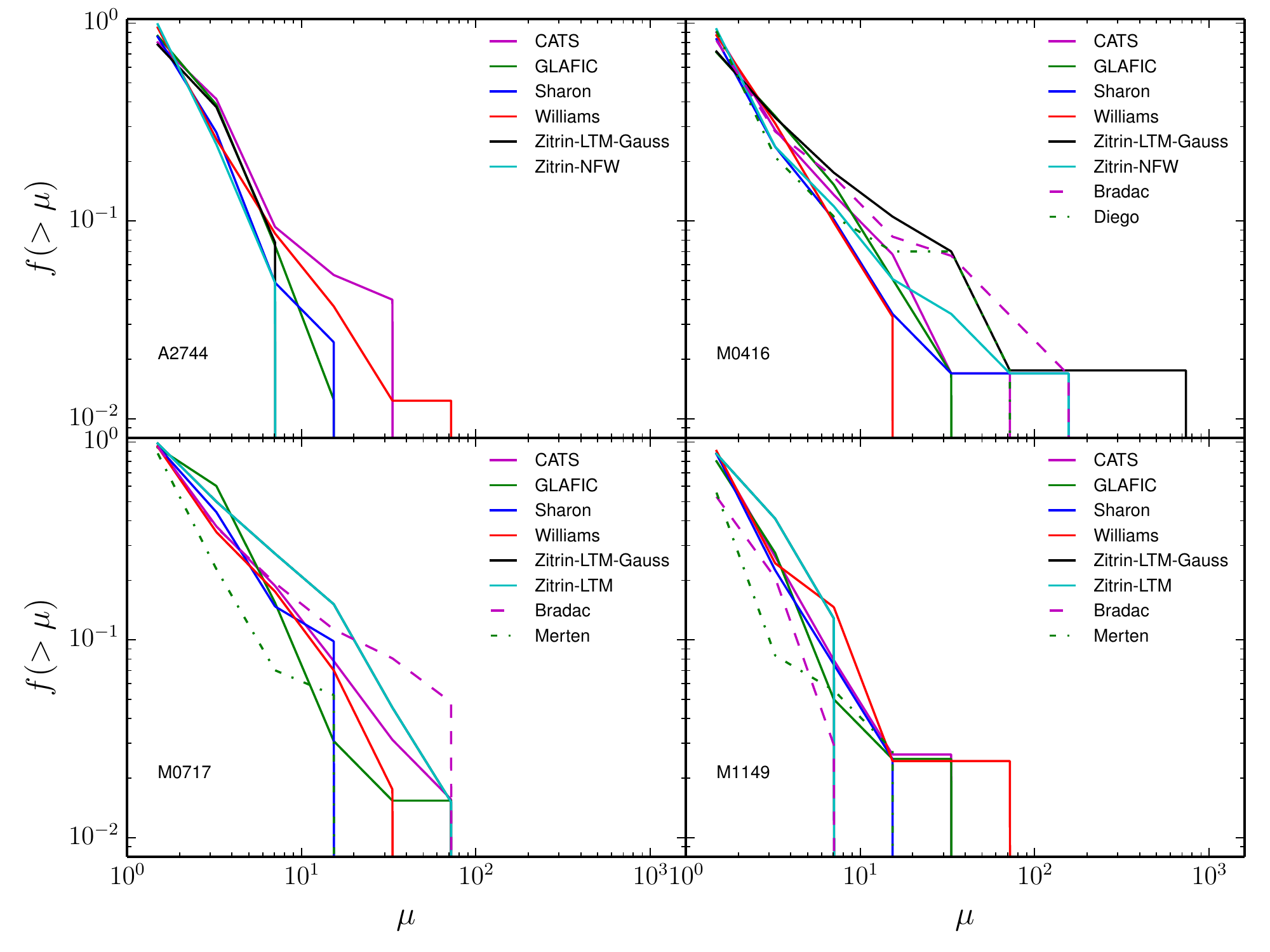}
\caption{
The distributions of the magnification factors of galaxies with photometric redshifts between redshift 5.0 and 9.5 in each cluster and $H_{\rm 160,int}$ in either of lensing models. 
}
\label{fig:magnifications}
}
\end{figure*}

Compared to C16a we update in this paper the A2744 and M0416 high-$z$ samples by exploiting the improved v3 lensing models now available, and we include in the analysis number counts from other two additional clusters, M0717 and M1149. In Tab. \ref{tb:clusters} we list the clusters and the corresponding lensing models used in this paper. 
In Fig. \ref{fig:magnifications} we plot the distributions of the magnification factors of our selected galaxy samples in each cluster, for our adopted lensing models.
In different lensing models an identified galaxy could have different magnifications, hence different demagnified magnitudes.
Therefore for a given cluster we can reconstruct different number counts (galaxy number per magnitude bin) when using different lensing models.  We make the median of these counts as our fiducial number counts.
In middle and bottom panels of Fig. \ref{fig:numbercount_obs} we show the number counts of the faint ($H_{\rm 160,int}>27.5$) galaxies 
with $5.0 < z < 7.0$ and $5.0<z<9.5$ in the fields of four FFs clusters
(we do not use the data of the parallel blank fields), as a function of $H_{\rm 160,int}$.

\begin{table*}
\caption{The FFs clusters and lensing models used in this paper.}
  \begin{tabular}{llllllllll}
    Cluster & Lensing model \\ \hline
    Abell 2744 (A2744) &GLAFIC v3; Sharon v3; Williams v3; Zitrin-LTM-Gauss v3;Zitrin-NFW v3; CATS v3.1 \\

    MACSJ0416.1-2403 (M0416)& GLAFIC v3; Sharon v3; Williams v3.1; Zitrin-LTM-Gauss v3; Zitrin-LTM v3; CATS v3.1; Brada\u{c} v3; Diego v3 \\
    MACSJ0717.5+3745 (M0717) &GLAFIC v3; Sharon v2; Williams v1; Zitrin-LTM-Gauss v1; Zitrin-LTM v1; CATS v1; Brada\u{c} v1; Merten v1 \\
    MACSJ1149.5+2223 (M1149) &GLAFIC v3; Sharon v2.1; Williams v1; Zitrin-LTM-Gauss v1; Zitrin-LTM v1; CATS v1; Brada\u{c} v1; Merten v1 \\
    \hline
  \end{tabular}
  \begin{tablenotes}
 \item Relevant references for lensing models listed in the table: GLAFIC: \citet{Kawamata2017,Kawamata2016,Ishigaki2015,Oguri2010}.
 Sharon: \citet{Johnson2014,Jullo2007}.
 Williams: \citet{Priewe2017,Sebesta2016,Grillo2015,Jauzac2014,Mohammed2014,Liesenborgs2006}.
 Zitrin: \citet{Zitrin2013,Zitrin2009}.
 CATS: \citet{Jauzac2015,Jauzac2014,Richard2014,Jauzac2012,Jullo2009}.
 Brada\u{c}: \citet{Hoag2016,Bradac2009,Bradac2005}. 
 Diego: \citet{Diego2015,Diego2007,Diego2005b,Diego2005a}.
 Merten: \citet{Merten2011,Merten2009}. 
 All models are available on the STSCI website.  
 \end{tablenotes}
  \label{tb:clusters}
\end{table*}

\subsection{An empirical description of the LF turn-over}\label{empirical}

Is there evidence of a ``turn-over'' in the faint-end of the high-$z$ galaxy LFs from the available FFs data? To investigate this problem we adopt the following reference LF model \--- a standard Schechter formula modulated by a term that rapidly drops when the absolute UV magnitude $M_{\rm UV}$ is much fainter than the turn-over magnitude $M_{\rm UV}^{\rm T}$, and rapidly approaches unity when $M_{\rm UV}\ll M_{\rm UV}^{\rm T}$:
\begin{align}
\Phi(M_{\rm UV},z)&=  0.4{\rm ln}(10) \Phi^* {\rm exp}[10^{-0.4(M_{\rm UV}-M_{\rm UV}^*)}] \nonumber \\
&\times 10^{-0.4(1+\alpha)(M_{\rm UV}-M_{\rm UV}^*)} 0.5[(1-{\rm erf}(M_{\rm UV}-M_{\rm UV}^{\rm T})],
\end{align}
where \textit{erf} is the error function. At the $M_{\rm UV}^{\rm T}$ the LF drops to half the value of a standard Schechter LF. In addition to the three redshift-dependent Schechter parameters $\Phi^*$, $M_{\rm UV}^*$, $\alpha$, we introduce here a new one parameter $M_{\rm UV}^{\rm T}$. The $M_{\rm UV}^*$ is mainly determined by observations of bright galaxies and by large volume galaxy surveys while it is unconstrained in our lensed samples of ultra-faint galaxies. For this reason, and to focus on the LF turn-over relevant parameters only,  we directly adopt the parameterization of $M_{\rm UV}^*$ from Sec. 5.1 of  \citet{Bouwens2015}:
 \begin{equation}
M_{\rm UV}^*=-20.95+0.01\times(z-6),
\end{equation}
discarding its uncertainties. We keep the $\Phi_*$ and $\alpha$ as free parameters that will be constrained together with $M_{\rm UV}^{\rm T}$ from the FFs data.

\subsection{A physically-motivated model of the high-$z$ galaxy LFs}\label{physical}

In Y16 we have developed a physically-motivated analytical model that describes the faint-end of the  high-$z$ galaxy LFs during the EoR. The model calibrates the ``star formation efficiency'' (defined as the star formation rate to halo dark matter mass ratio) - halo mass relation using the Schechter formula of observed LF at redshift $\sim5$, then computes the luminosity of a halo according to its mass and formation time at any redshifts (\citealt{Mason2015}, see also  \citealt{2010ApJ...714L.202T} and \citealt{2013ApJ...768L..37T}). Considering the probability distribution of a halo's formation time \citep{2007MNRAS.376..977G}, and the possibility of its star formation being quenched (if the circular velocity of this halo is smaller than a pre-assumed circular velocity criterion $v_c^*$ and it is located in ionized regions), the LF is then derived from halo mass function. In this model, the LF does not necessarily decrease monotonically at its faint end but have complex shapes, see Fig. 6  and Fig. 7 in Y16.  

The Y16 model has two free parameters, the escape fraction of ionizing photons, $f_{\rm esc}$, and the critical circular velocity, $v_c^*$. The galaxy number counts are sensitive to $v_c^*$ but less sensitive to $f_{\rm esc}$, therefore  we combine the number counts with the measured Thomson scattering optical depth to CMB photons, $\tau$, to obtain the joint constraints. 
 
\subsection{Statistical framework}

Here we summarise the procedure adopted to derive constraints on theoretical parameters from the  observed galaxy number counts. A more detailed discussion can be found in C16a.

The sample galaxies of each cluster in the specified redshift range are divided into $n_b$ bins according to their demagnified magnitudes. Suppose in the $i$th bin there are $N_{\rm obs}^i$ galaxies. For given luminosity function model with parameter set {\bf a}, we perform Monte Carlo simulations to calculate the probability to observe such number of galaxies in this bin, $p_1(N_{\rm obs}^i | {\bf a})$. 
In the Monte Carlo simulations, we include the completeness as a function size and magnitude of the image. The image size of each input galaxy is derived from its luminosity by using an intrinsic galaxy radius - luminosity relation given in \citet{2013ApJ...765...68H}. This relation is comparable with the relation in \citet{Bouwens2017-size,Bouwens2017-FFs}. We note that \citet{Kawamata2017} found a steeper relation slope for galaxies down to $M_{\rm UV}\sim-12.3$ in FFs, although at this moment we do not check the influence on our results. $p_1(N_{\rm obs}^i | {\bf a} )$ depends on both luminosity function models and lensing models. We use the mean probability of different lensing models (see their Eq. 3) except when comparing different lensing models.

We then build the following combined likelihood:
\begin{align}
L&= L_1\times L_2 \nonumber \\
&=\left[ \prod_i^{n_b} p_1(N_{\rm obs}^i | {\bf a})\right] \times L_2,
\label{eq_like}
\end{align}
where $L_1$ is the likelihood from our FFs observations, and $L_2$ is the likelihood of additional observations that can help to improve the constraints.

For the empirical model, we build $L_2$ from the constructed LF data points of wide blank fields at $z\sim6$,
\begin{equation}
L_2=\prod_j \frac{1}{\sqrt{2\pi \sigma^2_{\Phi,j} }}{\rm exp}\left[- \frac{ (\Phi_j-\Phi(\bf a))^2 }{2\sigma^2_{\Phi,j}}\right].
\end{equation}
Introducing this $L_2$ is necessary, because although the gravitational lensing surveys are deeper, usually they have smaller effective volume,  while the blank field surveys have large volume, thereby are helpful for reducing the uncertainties.

For the physically-motivated model, we build the $L_2$ from the measured CMB scattering optical depth, 
\begin{equation}
L_2({\bf a})=\frac{1}{\sqrt{2\pi \sigma^2_\tau}} {\rm exp}\left[-\frac{(  \tau_{\rm obs}  - \tau({\bf a}) )^2}{2\sigma^2_\tau}\right].
\end{equation}
The constraints on parameter set $\bf a$ are obtained by looking for the minimum of $\chi^2=-2{\rm log}(L)$ and its variations 
corresponding to different C.L. given by chi square distribution.

\section{Results}

\subsection{Is a LF turn-over observed at $z\sim6$ ?}

\begin{figure*}
\centering{
\includegraphics[scale=0.8]{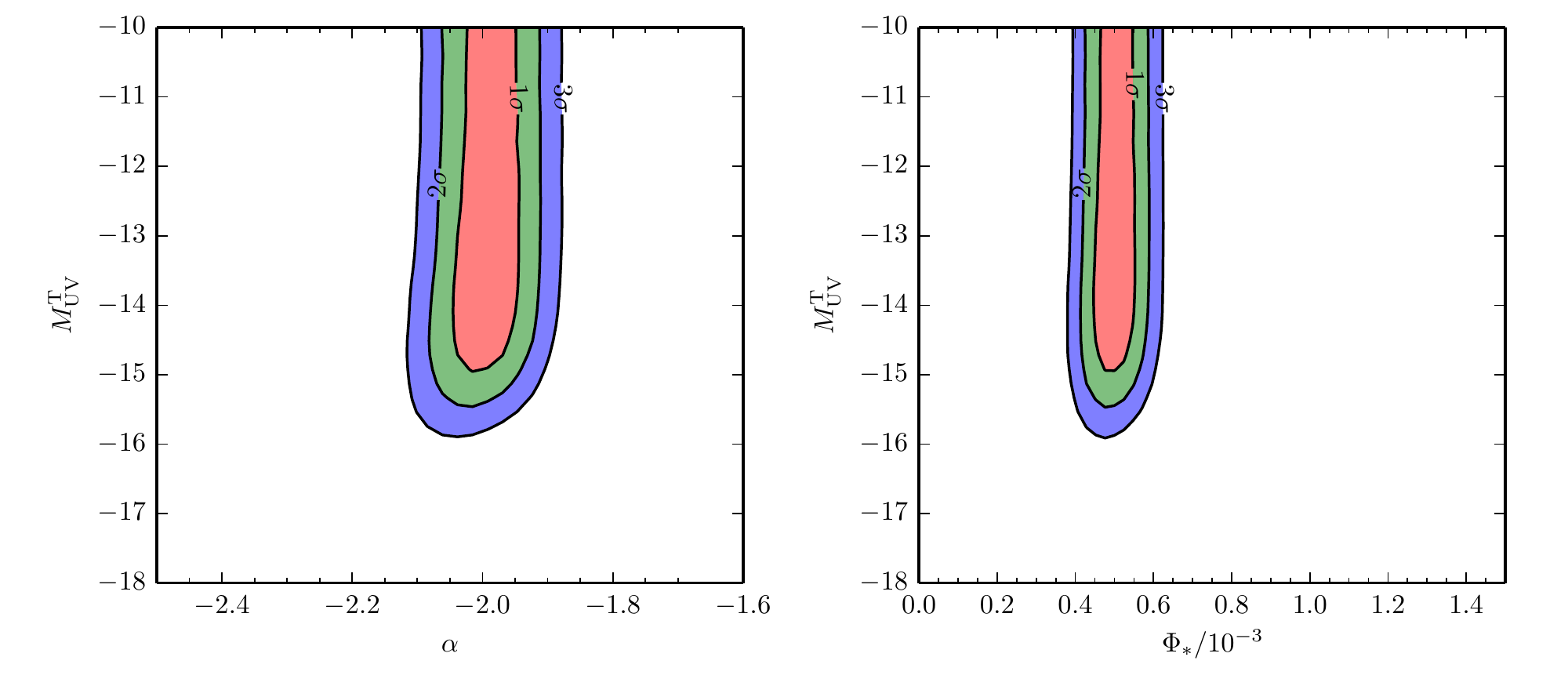}
\caption {The constraints on $\alpha$, $\Phi_*$ and $M_{\rm UV}^{\rm T}$  from the combination of all the four FFs clusters.  The panels are the $\alpha$-$M_{\rm UV}^{\rm T}$ and $\Phi_*$-$M_{\rm UV}^{\rm T}$ contour maps respectively. In each case the remaining parameter has been marginalized.
}
\label{fig:empirical_contour}
}
\end{figure*}

\begin{figure}
\centering{
\includegraphics[width=0.5\textwidth]{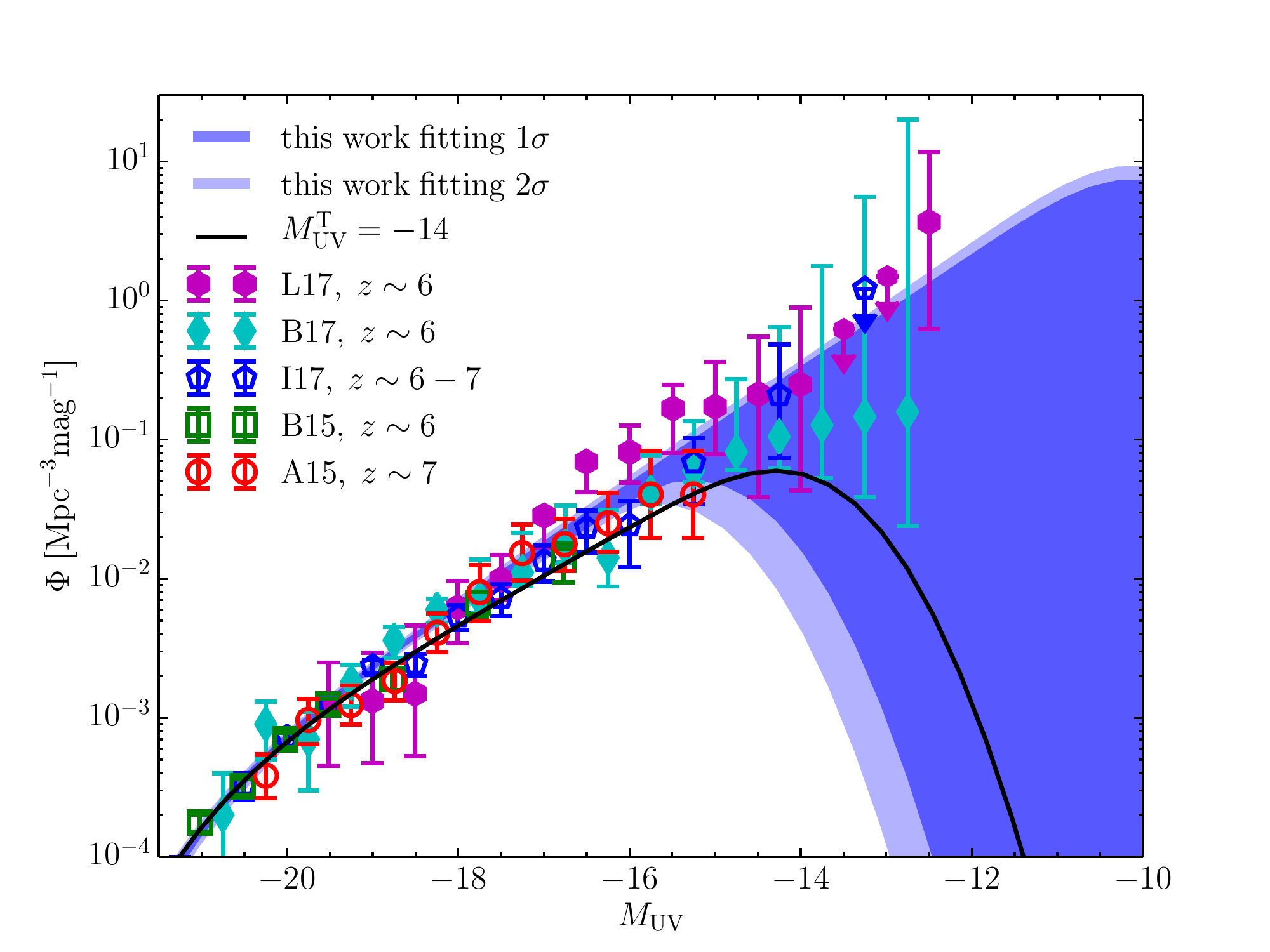}
\caption {The LF constrained in our work. We show the uncertainties within $1\sigma$ and $2\sigma$ C.L. As comparisons we 
plot the B15, A15, B17, L17 and I17 observations together, and a theoretical LF with $M_{\rm UV}^{\rm T}=-14$.}
\label{fig:empirical_LF}
}
\end{figure}

In this subsection we investigate the constraints on parameters in the empirical model described in Sec. \ref{empirical} at $z\sim6$ by analyzing galaxy samples with $5<z<7$ in ASTRODEEP catalogs. 

Using a collection of wide and deep blank field {\it HST} surveys data, including the CANDELS, HUDF09, HUDF12, ERS and BoRG/HIPPIES fields, \citet{Bouwens2015} (B15) have  constructed the LFs from $z\sim4$ to $z\sim10$. We use their stepwise maximum likelihood determination of the $z\sim6$ data points to build the $L_2$ (see Table 5 of B15).  
We vary $\Phi_*$ in the range $\Phi_*\le1.5\times10^{-3}$ Mpc$^{-3}$, $\alpha$ in the range $-2.5\le \alpha\le-1.6$ and $M_{\rm UV}^{\rm T}$ in the range $-18\le M_{\rm UV}^{\rm T}\le-10$.

The constraints on empirical model parameters are shown in Fig. \ref{fig:empirical_contour}. In the 2D $M_{\rm UV}^{\rm T}$-$\alpha$ contour map, the $\Phi_*$ has been marginalized, and in the $M_{\rm UV}^{\rm T}$-$\Phi_*$ contour map the $\alpha$ has been marginalized. We can see that the upper boundary of $M_{\rm UV}^{\rm T}$ is always open. To obtain the final constraints we marginalize both $\Phi_*$ and $\alpha$, 
and we have $M_{\rm UV}^{\rm T}\gsim-14.6$ at $1\sigma$ C.L. and $M_{\rm UV}^{\rm T}\gsim-15.2$  at $2\sigma$ C.L.. 
Still, we only find the lower boundary of $M_{\rm UV}^{\rm T}$, the upper boundary is open. This implies that no evidence is found in the existing data for the four FFs clusters of a LF turn-over at $z\approx 6$. We summarise the constraints in Tab. \ref{tb:constraints}.  
We have tested that in Eq. (\ref{eq_like}) if we use the $L_2$ derived from \citet{Finkelstein2015} LF at $z\sim6$ we obtained quite similar results on $M_{\rm UV}^{\rm T}$.
If we remove $L_2$, i.e. using only the FFs data, while we restrict $\Phi_*$, $\alpha$  and $M_{\rm UV}^{\rm T}$ vary in range specified in last paragraph, we obtain $M_{\rm UV}^{\rm T}\gsim-15.2$ and $M_{\rm UV}^{\rm T}\gsim-15.9$ at 1$\sigma$ and  2$\sigma$ C.L. respectively.

We plot the LF corresponding to the constraints at $z\sim6$ in Fig. \ref{fig:empirical_LF} by curves and filled regions. As a reference and consistency check, we also plot the B15 LF data, and the  LF data constructed from A2744, M0416 and M0717 and their corresponding parallel blank filed in  A15 at $z\sim7$, which is one of the deepest LFs and consistent with other LFs in the overlap magnitude range.  
Moreover, the L17, B17 and I17 LFs are also plotted in Fig. \ref{fig:empirical_LF}.

Before making comparisons between B17 and our results, it is necessary to clarify a dissimilarity between the definition of the ``turn-over magnitude'' between B17 and our work. In B17, the turn-over magnitude is the absolute magnitude at which the LF's derivative is zero, while in our work it is defined as the absolute magnitude where the LF decreases to half of the Schechter LF. Moreover, their LF-end is modulated by a term $10^{-0.4\delta(M_{\rm UV}+16)^2}$ which could decrease gently even when $M_{\rm UV}$ is higher than the turn-over magnitude, depending on $\delta$.  However we assume the modulation term $0.5[1-{\rm erf}(M_{\rm UV}-M_{\rm UV}^{\rm T})]$, at $M_{\rm UV}> M_{\rm UV}^{\rm T}$ the LF drops rapidly. For the above reason, we do not directly compare our $M_{\rm UV}^{\rm T}$ with B17. Instead, we plot our constrained LF together with the LF constructed in B17, see Fig. \ref{fig:empirical_LF}. From this figure, the B17 LF constraint could be approximately translated into $M_{\rm UV}^{\rm T}\gsim-14$ of our model, 
a bit deeper than what we found in our work,  $M_{\rm UV}^{\rm T}\gsim-14.6$. 
This difference might be due to the different methodologies adopted to take into account systematic effects. We use the median of the number counts from different lensing models as the true number count, while B17 incorporates systematics estimated by the difference between the various models. 
As a check, in our work if we drop the galaxies with magnification factors $>100$ in observations we obtain constraints $M_{\rm UV}^{\rm T}\gsim-14.8$ ($-15.4$) at $1\sigma$ ($2\sigma$) C.L., quite similar to the model without dropping galaxies.

The L17 constraint on LF turn-over is deeper than ours, i.e. no turn-over is seen until $M_{\rm UV}\sim-12.5$ at $z\sim6$ in their work. 
The difference between our results and L17 ones can be due to the different methodologies adopted: a) They build the source catalog and subtract the intra-cluster light in a very different way than in our case. b) They have assumed different galaxy size distributions.
c) In our case, we first construct the number count for each lensing model independently, then take the median of the number counts reconstructed from different lensing models; while in L17, for the image of each galaxy they take the flux-weighted magnification of different lensing models, then construct the LF.

\begin{figure*}
\centering{
\subfigure{\includegraphics[scale=0.7]{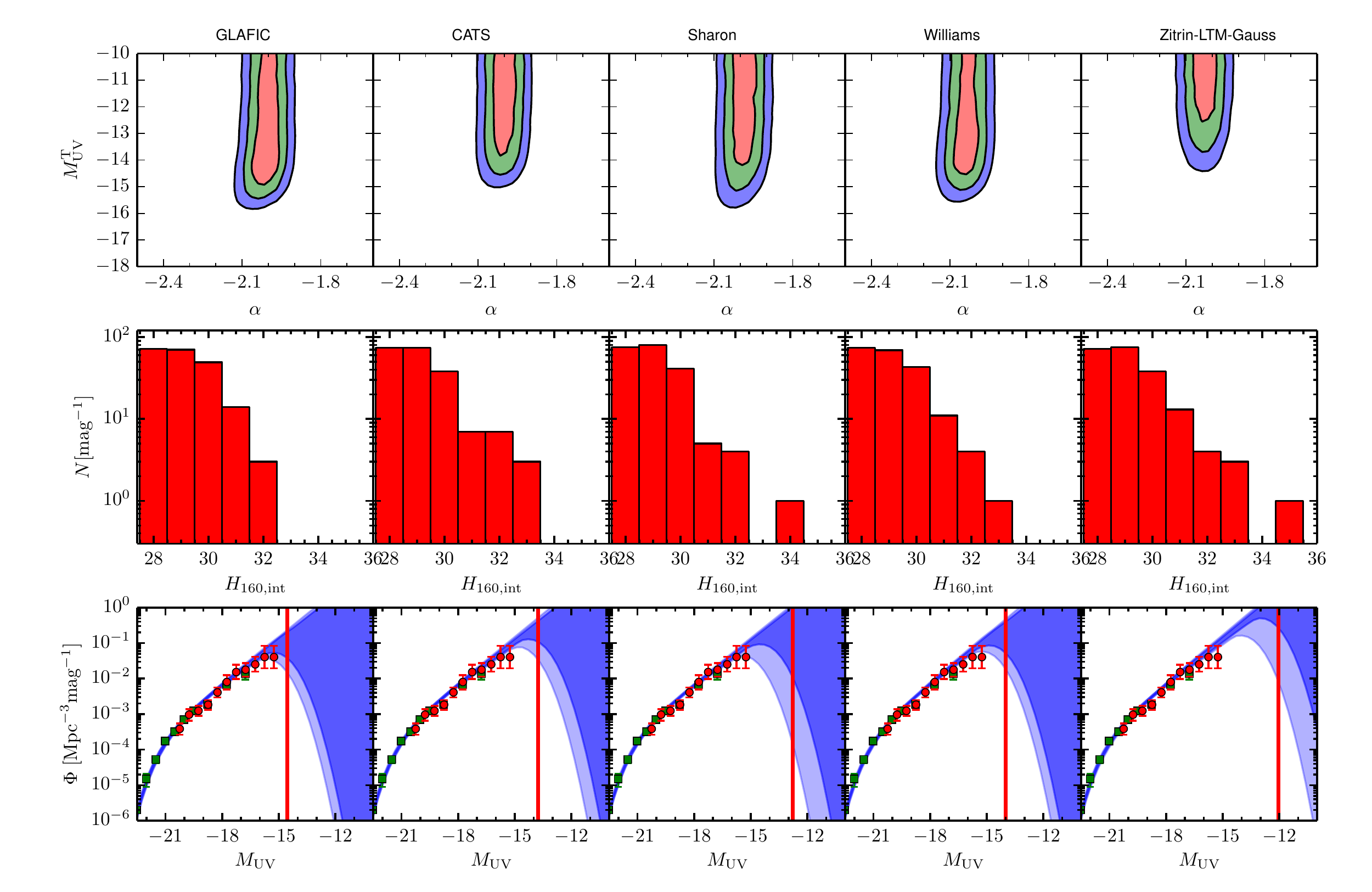}}
  \caption {{\it Top panels:} The constraints on parameters $\alpha$ and $M_{\rm UV}^{\rm T}$ from five individual lensing models (marked in the panel), ``GLAFIC", ``CATS", ``Sharon", ``Williams" and ``Zitrin-LTM-Gauss" from the four FFs clusters data in $z\sim5-7$. {\it Middle panels:} The sum of the number counts for the four FFs clusters in $z\sim5-7$ for the five lensing models. {\it Bottom panels:} 
The LFs corresponding to constraints at 1$\sigma$ C.L. (regions filled by deeper colors) and 2$\sigma$ C.L. (regions filled by lighter colors). To have clear panels we now only plot the B15 (squares) and A15 (circles) data.    
In the LFs panels we mark the absolute UV magnitudes of the faintest galaxies identified using each lensing model by vertical lines.
}
\label{fig:empirical_lensing}
}
\end{figure*}

We now investigate the systematic differences between the various lensing models. In Fig. \ref{fig:empirical_lensing} we show the  
$\alpha$-$M_{\rm UV}^{\rm T}$ constraints by using only one lensing model each time, ignoring version discrepancies.  Each of ``GLAFIC", ``CATS", ``Sharon", ``Williams" and ``Zitrin-LTM-Gauss" is shown by one column in Fig. \ref{fig:empirical_lensing}. We choose these five lensing models because they are available for all the four clusters; and at least for A2744 and M0416 the versions are equal or later than v3.0.  For each lensing model from top to bottom the panels are $\alpha$-$M_{\rm UV}^{\rm T}$ constraints, the sum of the number counts for the four FFs clusters and the LFs corresponding to the constraints (see Tab. \ref{tb:constraints}).

Indeed, the discrepancies between lensing models are rather evident, especially for the $M_{\rm UV}^{\rm T}$ boundaries. This is because these lensing models use different mass distribution and observations as constraint inputs;
as a results,  although the number counts (the middle panels of Fig. \ref{fig:empirical_lensing}) are basically consistent with each at $H_{\rm \rm 160,int}\lsim 32$, at the faintest end they are rather different from each other.
Detailed investigations about the systematics among lensing models could be found in \citet{Priewe2017,FFs_systematic2017,lensing_model_comparison2017} and the references of each lensing models listed below Tab. \ref{tb:clusters}.
In all the cases, the upper boundaries are open, implying that no turn-over is apparent.

In bottom panels of Fig. \ref{fig:empirical_lensing} we also plot the absolute magnitudes of the faintest galaxies in each lensing model  by vertical lines. Usually the $M_{\rm UV}^{\rm T}$ constraints are shallower than these faintest magnitudes. We check the influence of the faintest galaxies on the $M_{\rm UV}^{\rm T}$ constraints. 
We find that for all lensing models except Williams discussed in Fig. \ref{fig:empirical_lensing}, the faintest galaxies (referring to the demagnified magnitudes) are in the M0416 field. For the Williams model the faintest galaxy is in the M0717 field.

For the GLAFIC and CATS lensing models, the faintest galaxy is the same one whose observed apparent magnitude $H_{\rm 160}=28.1$, and demagnified magnitudes $H_{\rm 160,int}=32.2$ and 32.9 in these two lensing models respectively. 
For Sharon, Williams and  Zitrin-LTM-Gauss lensing model, the faintest galaxies have $H_{160}=28.0$, 28.6 and 26.7, 
and  $H_{\rm 160,int}=33.9$, 32.5 and 34.6 respectively. An investigation of the influence by the photometric errors is given in Appendix \ref{photo_errors}.

Although we have checked all the galaxies  one-by-one visually and do not find any reason to consider the above faint galaxies passed our checkup spurious objects, we still make a test to check their influences. 
In case including (removing) them in samples, we obtain the 2$\sigma$ C.L. constraints: $M_{\rm UV}^T\gsim$-15.2 (-15.3), -14.3 (-14.5), -14.7 (-15.5), -14.9 (-15.2), and -13.2 (-14.1) for lensing models GLAFIC, CATS, Sharon, Williams and Zitrin-LTM-Gauss respectively.  The changes on the Sharon and Zitrin-LTM-Gauss model are most obvious, almost up to 1 magnitude.

\begin{table*}
\centering{
\caption{Constraints on $M_{\rm UV}^{\rm T}$ and $v_c^*$, and the halo mass and absolute UV magnitude corresponding to $v_c^*$ constraints}

\begin{tabular}{llllllll}
        
     &  & ALL & GLAFIC & CATS & Sharon & Williams & Zitrin-LTM-Gauss \\ \hline\hline
       
  \multirow{2}{*}{$M_{\rm UV}^{\rm T}$}  & $1\sigma$  &$ \gsim-14.6$ & $\gsim-14.6$ &   $\gsim-12.9$ &   $\gsim-13.7$ & $\gsim-14.3$ & $\gsim-11.8$  \\   
   
                                                                & $2\sigma$  &$ \gsim-15.2$ & $ \gsim-15.2$ &   $ \gsim-14.3$ & $ \gsim-14.7$ & $ \gsim-14.9$ &  $ \gsim-13.2$ \\

\hline

 \multirow{2}{*}{$v_c^*/{\rm km s^{-1}}$}  & $1\sigma$  &$ \lsim50$  & $ \lsim48 $ &   $ \lsim40$   &    $ \lsim45$ &  $ \lsim45$    &   $ \lsim34$   \\
    
                                                                 &$2\sigma$   &$ \lsim59$   & $ \lsim56$   &   $ \lsim49$   &   $ \lsim56$  &  $ \lsim54$  &  $\lsim45$   \\         
\hline

 \multirow{2}{*}{$M_h/M_\odot(z=5)$}  & $1\sigma$  &$ \lsim5.6\times10^9$  & $ \lsim4.9\times10^9 $ &   $ \lsim2.9\times10^9$   &    $ \lsim4.1\times10^9$ &  $ \lsim4.1\times10^9$    &   $ \lsim1.8\times10^9$   \\
    
                                                             &$2\sigma$&$ \lsim9.2\times10^9$  & $ \lsim7.9\times10^9 $ &   $ \lsim5.3\times10^9$   
  &   $ \lsim7.9\times10^9$ &  $\lsim7.0\times10^9$    &  $ \lsim4.1\times10^9$   \\
  
  \hline
  
   \multirow{2}{*}{$M_{\rm UV}(z=5)$}  & $1\sigma$  &$ \gsim-14.2$  & $ \gsim-14.0 $ &   $ \gsim-13.2$   &    $ \gsim-13.7$ &  $ \gsim-13.7$    &   $ \gsim-12.4$   \\
    
                                                            &$2\sigma$&$ \gsim-15.0$  & $ \gsim-14.8 $ &   $ \gsim-14.1$   
  &   $ \gsim-14.8$ &  $\gsim-14.6$    &  $ \gsim-13.7$   \\

  \hline
  
  \multirow{2}{*}{$M_h/M_\odot(z=9.5)$}  & $1\sigma$  &$ \lsim2.4\times10^9$  & $ \lsim2.1\times10^9 $ &   $ \lsim1.2\times10^9$   &    $ \lsim1.8\times10^9$ &  $ \lsim1.8\times10^9$    &   $ \lsim7.6\times10^8$   \\
  
                                                                & $2\sigma$  &$ \lsim4.0\times10^9$  & $ \lsim3.4\times10^9 $ &   $ \lsim2.3\times10^9$   &    $ \lsim3.4\times10^9$ &  $ \lsim3.0\times10^9$    &   $ \lsim1.8\times10^9$   \\

\hline

\multirow{2}{*}{$M_{\rm UV}(z=9.5)$}  & $1\sigma$  &$ \gsim-14.0$  & $ \gsim-13.8 $ &   $ \gsim-13.0$   &    $ \gsim-13.6$ &  $ \gsim-13.6$    &   $ \gsim-12.3$   \\
    
                                                            &$2\sigma$&$ \gsim-14.8$  & $ \gsim-14.6$ &   $ \gsim-14.0$   
  &   $ \gsim-14.6$ &  $\gsim-14.4$    &  $ \gsim-13.6$   \\
  
  \hline

\end{tabular}
    
}
         
 \label{tb:constraints}

\end{table*}

\subsection{Constraints on $v_c^*$}

\begin{figure}
\centering{
\subfigure{\includegraphics[scale=0.4]{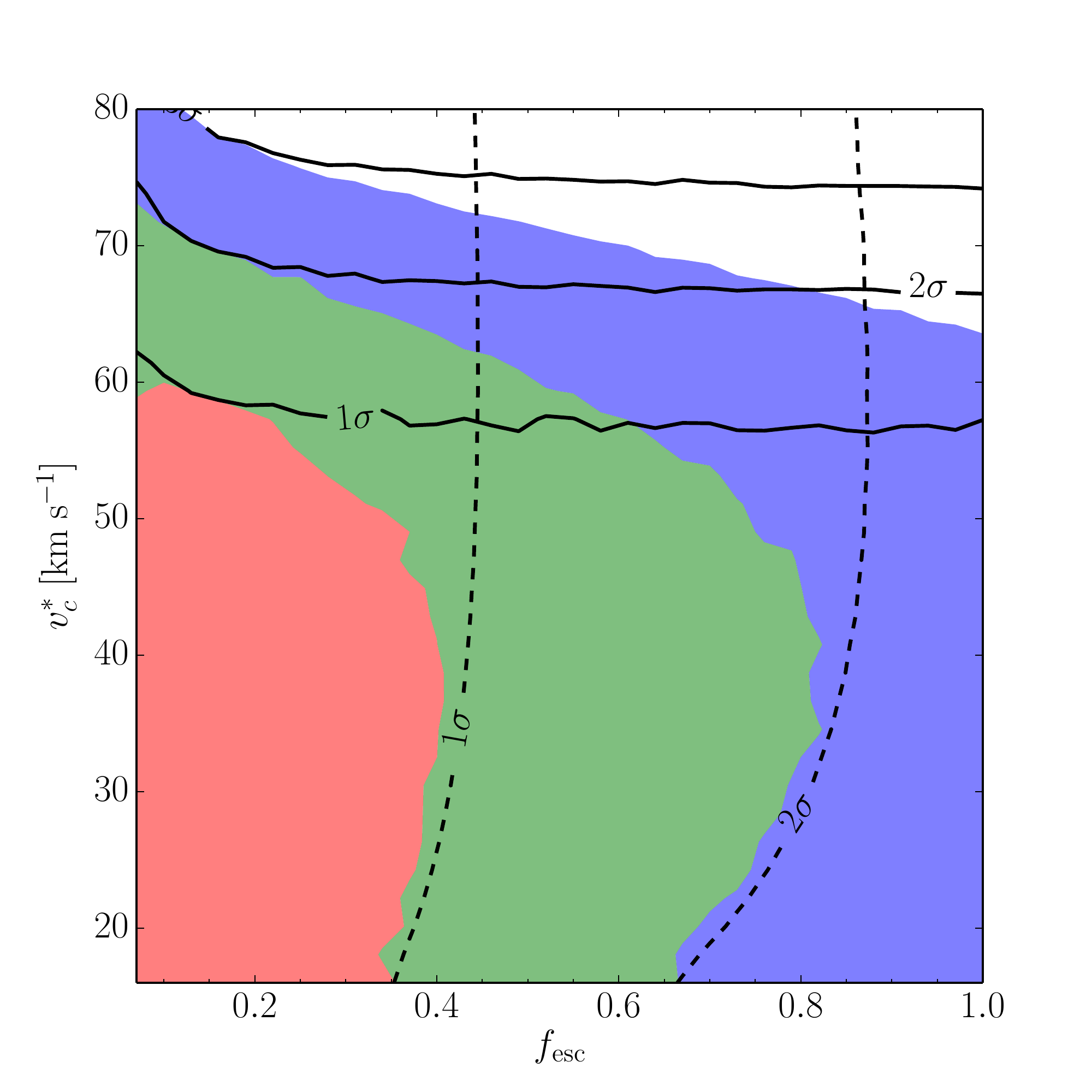}}
\subfigure{\includegraphics[scale=0.4]{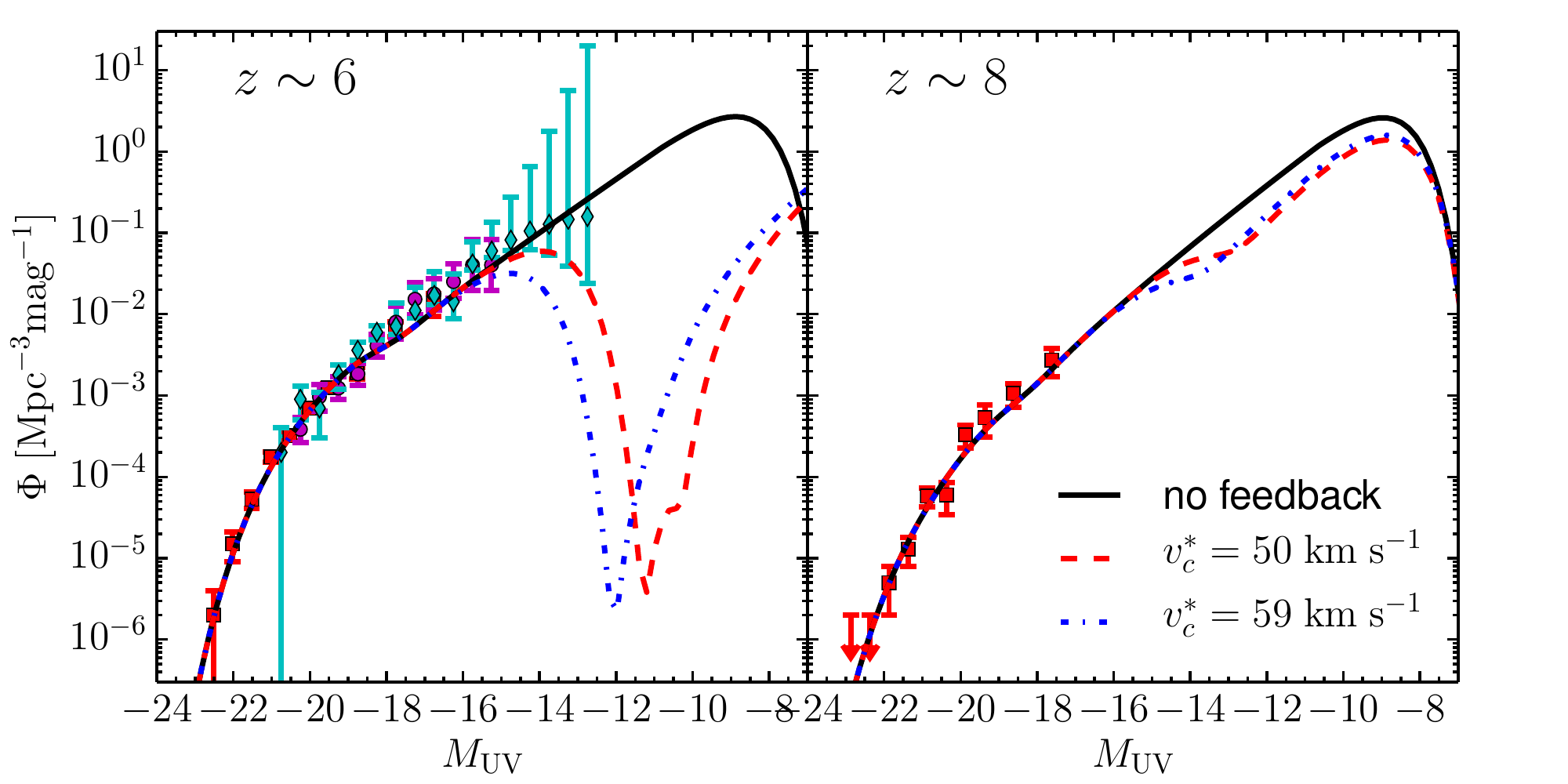}}
\caption {
{\it Top:} 
The constraints on $f_{\rm esc}$ and $v_c^*$. We obtain them by using the combination of FFs galaxy number counts and the {\it Planck}2016 CMB scattering optical depth $\tau=0.058\pm0.012$  \citep{Planck2016-reionization}. Solid contour lines refer to the constraints obtained from galaxy number counts only,  dashed lines refer to from the CMB only,  and filled regions refer to constraints from their combination.
{\it Bottom:} The LFs that correspond to no feedback, $v_c^*=50$ and $59$ km s$^{-1}$ models at $z\sim6$ and 8.  Filled squares refer to the B15 data at $z\sim6$ and 8; filled circles refer to A15 data at $z\sim7$; and filled diamonds refer to B17 data at $z\sim6$. We always use $f_{\rm esc}=0.15$ for theoretical LFs in this panel. 
}
\label{fig:physical}
}
\end{figure}

We then investigate the constraints we can put on the physically-motivated model. Since in this model both $f_{\rm esc}$ and $v_c^*$ are redshift independent parameters, we use all the data in $z=5.0-9.5$.
In top panel of Fig. \ref{fig:physical} we show the constraints on $f_{\rm esc}$ and $v_c^*$, using the combination of galaxy number counts in the FFs fields and the latest {\it Planck}2016 Thomson scattering optical depth to CMB photons: $\tau=0.058\pm0.012$ \citep{Planck2016-reionization}.  
Compared with C16a, the smaller $\tau$ helps us to obtain tighter constraints on $f_{\rm esc}$, say $f_{\rm esc}\lsim57\%$ ($2\sigma$ C.L.) after marginalizing $v_c^*$. 
When marginalizing $f_{\rm esc}$ we find $v_c^*\lsim 59 $ km s$^{-1}$ ($2\sigma$ C.L.), which corresponds to a halo mass $M_h\lsim4.0\times10^9~M_\odot$ and $9.2\times10^9~M_\odot$ at $z=9.5$ and 5 respectively. 
Given the halo mass, using the star formation efficiency - halo mass relation constructed in Y16, and the halo assembly history, we can derive its mean luminosity. We therefore translate the halo mass constraints into absolute UV magnitude constraints, 
$M_{\rm UV}\gsim -14.8$ and $-15.0$. They are slightly tighter than  those reported in C16a (see Fig. 3 there). 
Additional constraints are listed in Tab. \ref{tb:constraints}, where we present both 1$\sigma$ and 2$\sigma$ constraints. The LFs at $z\sim6$ and 8 in Y16 model corresponding to no feedback, $v_c^*=50$ and $59$ km s$^{-1}$ for $f_{\rm esc}=0.15$ are shown in bottom panels of Fig. \ref{fig:physical}.

We also find that different clusters do not contribute equally to the final constraint. 
If we respectively {\it remove} one of A2744, M0416, M0717 and M1149 each time, 
we obtain $v_c^*\lsim$65, 61, 62 and 58 km s$^{-1}$ (all at 2$\sigma$ C.L.) respectively. 
As seen in Fig. \ref{fig:numbercount_obs}, the M1149 has less contribution to the number counts in the faintest magnitude bin,  therefore have smaller influence in the final constraint.

We also investigate the discrepancies between different lensing models in this case. When using one lensing model at a time, as mentioned in the last subsection, we obtain the $2\sigma$ C.L. constraints: $v_c^*\lsim$56 (GLAFIC), 49 (CATS), 56 (Sharon), 54 (Williams) and 45 (Zitrin-LTM-Gauss) km s$^{-1}$ respectively, see Tab. \ref{tb:constraints}.

\section{Conclusions}
 
We investigated the LF of galaxies in the reionization epoch at low luminosities under the explicit assumption that any deviation from a pure Schechter LF at faint magnitudes is imprinted by feedback effects during reionization itself. We considered two LF models, and obtained constraints on their parameters from the observed high-$z$ ultra-faint galaxy number counts in four FFs gravitational lensing cluster fields. We first test an empirical model where the standard Schechter formula is modulated by the suppressing term $0.5[1-{\rm erf} (M_{\rm UV}-M_{\rm UV}^{\rm T})]$. The LF is unchanged when $M_{\rm UV}\ll M_{\rm UV}^{\rm T}$, and drops rapidly when $M_{\rm UV}\gg M_{\rm UV}^{\rm T}$. Secondly, we consider the physically-motivated model proposed by Y16 and analysed in C16a. In this model the star formation is quenched in halos with circular velocity smaller than $v_c^*$,  during and after the EoR, as long as they are located in ionized regions. As a result, the LF has complex behavior at low luminosities.

We used the photometric catalogs and redshifts of the four FFs clusters A2744, M0416, M0717 and M1149 provided by the ASTRODEEP collaboration. The first two clusters have already been analyzed in a previous work C16a, therefore in this paper we only considered the lensing models with versions later than 3.0, which were not adopted in C16a. For other two clusters, we used all available lensing models, and where multiple versions are available we adopted the latest ones.

For the empirical model, at 1$\sigma$ (2$\sigma$) C.L. we have obtained constraints $M_{\rm UV}^{\rm T}\gsim-14.6$~($M_{\rm UV}^{\rm T}\gsim-15.2$) at $z\sim6$. 
We therefore concluded that we have not yet confirmed the LF turn-over in the data of these four FFs clusters.

For the physically-motivated model we obtained $v_c^*\lsim59$ km s$^{-1}$ at 2$\sigma$ C.L., corresponding to absolute UV magnitude $-15.0$ at $z=9.5$ and $-14.8$ at $z=5$.
Considering the discrepancies between different lensing models, 
we have $v_c^*\lsim45 \--  59$ km s$^{-1}$. 
In all the cases considered, we have not found the lower limit for $v_c^*$.

All the numerical results of both the empirical model and the physically-motivated model are listed in Tab. \ref{tb:constraints}, and in the physically-motivated model we have translated the $v_c^*$ constraints into the $M_{\rm UV}$ constraints at $z=5$ and $z=9.5$ respectively. Although the constraints on $v_c^*$ can be translated into constraints on $M_{\rm UV}^{\rm T}$ through the luminosity - halo mass relations, we remind that in empirical model the constraints are purely from the galaxy surveys, while in the physically-motivated model the constraints are from both the galaxy surveys and the CMB scattering optical depth. 
In spite of this, the results of these models  are considered consistent in the fiducial case (ALL model in Tab. \ref{tb:constraints}): 
e.g. $M_{\rm UV}^{\rm T}\gsim-14.6$ vs. $M_{\rm UV}(z=5)\gsim-14.2$.

Thanks to the combined power of gravitational lensing and deep {\it HST} multi-band imaging we are just starting to observe the faintest galaxy populations likely responsible for reionization. The present analysis and similar ones in the past have not yet found significant evidence of the presence of feedback effects suppressing the formation of galaxies at faint UV magnitudes. This is likely due to the uncertainties and systematics involved in lensing models and in the selection and characterization of distant, faint sources.  In this respect, the completion of the FFs survey, and improvements in lensing model accuracy as well as high redshift sample selection enabled by future {\it JWST} photometric and spectroscopic observations will be crucial for improving our understanding of reionization.

\acknowledgments

 We thank the anonymous referee for the useful suggestions helpful for improving the paper. 
This work utilizes gravitational lensing models produced by PIs Brada\u{c}, Natarajan \& Kneib (CATS), Merten \& Zitrin, Sharon, and Williams, and the GLAFIC and Diego groups. This lens modeling was partially funded by the HST Frontier Fields program conducted by STScI. STScI is operated by the Association of Universities for Research in Astronomy, Inc. under NASA contract NAS 5-26555. The lens models were obtained from the Mikulski Archive for Space Telescopes (MAST). 
BY acknowledges the support of  the CAS Pioneer Hundred Talents (Young Talents) program, the NSFC grant 11653003, the NSFC-CAS  joint fund for space scientific satellites No. U1738125, and the NSFC-ISF joint research program No. 11761141012.
RA acknowledges support from the ERC Advanced Grant 695671 ÒQUENCHÓ. M.J.M.~acknowledges the support of  the National Science Centre, Poland through the POLONEZ grant 2015/19/P/ST9/04010. This project has received funding from the European Union's Horizon 2020 research and innovation programme under the Marie Sk{\l}odowska-Curie grant agreement No. 665778.

\bibliographystyle{yahapj}
\bibliography{refe}

\clearpage

\begin{appendix}

\section{The list of our selected galaxy samples}
\label{gal_list}

\setcounter{table}{0}
\renewcommand{\thetable}{A\arabic{table}}

\begin{ThreePartTable} 
\begin{longtable}{lll | lll | lll | llll }   
\caption{Our selected galaxies in ASTRODEEP catalogs. The SEDs, cutouts and all ancillary information could be found on the ASTRODEEP CDS interface at   \url{http://astrodeep.u-strasbg.fr/ff/}. } \\
\toprule
\multicolumn{3}{c}{A2744} & \multicolumn{3}{c}{M0416}  &  \multicolumn{3}{c}{M0717} &  \multicolumn{3}{c}{M1149} \\
\cmidrule(lr){1-3}\cmidrule(lr){4-6} \cmidrule(lr){7-9} \cmidrule(lr){10-12} \\
ID  & $H_{160}$ & $z$  &  ID  & $H_{160}$ & $z$    & ID  & $H_{160}$ & $z$ & ID  & $H_{160}$ & $z$ \\
 \midrule
54& 27.81$\pm$  0.15&  5.10$\pm$  0.11&      73& 28.03$\pm$  0.15&  6.07$\pm$  0.07&      69& 27.09$\pm$  0.12&  5.18$\pm$  0.09&     354& 28.03$\pm$  0.15&  5.66$\pm$  0.14&\\
      62& 25.18$\pm$  0.05&  6.13$\pm$  0.03&     132& 28.11$\pm$  0.26&  5.03$\pm$  0.08&      75& 28.56$\pm$  0.22&  5.19$\pm$  0.47&     362& 27.14$\pm$  0.09&  5.61$\pm$  0.17&\\
      67& 27.17$\pm$  0.12&  5.54$\pm$  0.07&     141$^*$& 27.77$\pm$  0.13&  6.68$\pm$  0.99&      82& 26.77$\pm$  0.08&  6.05$\pm$  0.05&     396& 28.35$\pm$  0.18&  6.92$\pm$  1.01&\\
      73& 26.50$\pm$  0.05&  6.85$\pm$  0.04&     143$^*$& 26.97$\pm$  0.08&  6.55$\pm$  1.00&      96& 27.44$\pm$  0.19&  7.90$\pm$  0.04&     402& 28.17$\pm$  0.19&  6.38$\pm$  0.08&\\
     145& 27.08$\pm$  0.13&  5.87$\pm$  0.30&     158& 28.00$\pm$  0.24&  8.27$\pm$  0.10&     151& 27.65$\pm$  0.18&  5.43$\pm$  0.05&     433& 27.75$\pm$  0.13&  6.16$\pm$  0.02&\\
     189& 28.01$\pm$  0.14&  5.68$\pm$  0.15&     201$^*$& 28.70$\pm$  0.21&  5.70$\pm$  2.45&     165& 27.52$\pm$  0.17&  5.49$\pm$  3.24&     448& 28.66$\pm$  0.26&  5.05$\pm$  0.19&\\
     203& 28.59$\pm$  0.19&  5.53$\pm$  0.03&     220& 27.74$\pm$  0.15&  5.82$\pm$  0.06&     222$^*$& 27.53$\pm$  0.26&  5.05$\pm$  0.85&     531& 27.92$\pm$  0.14&  6.56$\pm$  0.13&\\
     222& 27.23$\pm$  0.11&  5.02$\pm$  0.03&     246& 28.81$\pm$  0.40&  6.41$\pm$  0.02&     248$^*$& 27.73$\pm$  0.16&  5.42$\pm$  0.53&     546& 28.22$\pm$  0.26&  5.01$\pm$  0.03&\\
     263& 27.08$\pm$  0.12&  5.25$\pm$  0.09&     247$^*$& 27.22$\pm$  0.10&  7.13$\pm$  0.11&     272& 28.29$\pm$  0.20&  5.66$\pm$  0.12&     574& 28.01$\pm$  0.20&  5.21$\pm$  2.12&\\
     292$^*$& 27.59$\pm$  0.10&  5.38$\pm$  0.03&     265& 27.58$\pm$  0.21&  5.24$\pm$  2.00&     336& 27.10$\pm$  0.10&  6.10$\pm$  0.05&     708& 28.10$\pm$  0.19&  5.14$\pm$  0.11&\\
     321& 28.83$\pm$  0.19&  5.71$\pm$  0.03&     286$^*$& 28.20$\pm$  0.17&  8.14$\pm$  0.18&     356& 28.45$\pm$  0.23&  6.16$\pm$  0.12&     905& 25.93$\pm$  0.03&  5.05$\pm$  0.10&\\
     345$^*$& 28.46$\pm$  0.17&  5.35$\pm$  0.13&     354& 28.41$\pm$  0.23&  5.24$\pm$  0.05&     361& 26.17$\pm$  0.10&  6.45$\pm$  0.03&     942$^*$& 26.90$\pm$  0.09&  6.25$\pm$  3.93&\\
     379& 28.15$\pm$  0.16&  6.61$\pm$  0.21&     355& 27.87$\pm$  0.17&  6.24$\pm$  0.03&     374$^*$& 27.37$\pm$  0.17&  6.25$\pm$  0.08&     945& 27.92$\pm$  0.13&  6.68$\pm$  0.09&\\
     389& 27.44$\pm$  0.10&  5.36$\pm$  0.08&     465& 27.40$\pm$  0.11&  5.33$\pm$  0.09&     392$^*$& 27.54$\pm$  0.24&  6.03$\pm$  0.20&    1144& 27.96$\pm$  0.14&  5.41$\pm$  0.19&\\
     394& 28.40$\pm$  0.17&  6.60$\pm$  0.06&     513$^*$& 28.12$\pm$  0.21&  6.79$\pm$  2.58&     471& 26.68$\pm$  0.08&  6.53$\pm$  0.03&    1180& 27.31$\pm$  0.14&  7.17$\pm$  0.14&\\
     397& 27.17$\pm$  0.10&  6.31$\pm$  0.11&     524& 28.11$\pm$  0.15&  5.20$\pm$  0.07&     510& 26.87$\pm$  0.08&  6.46$\pm$  0.03&    1226& 28.35$\pm$  0.17&  9.14$\pm$  3.51&\\
     409& 28.32$\pm$  0.19&  7.66$\pm$  0.01&     637$^*$& 27.69$\pm$  0.13&  6.41$\pm$  0.20&     511& 27.36$\pm$  0.11&  5.20$\pm$  0.06&    1243& 27.44$\pm$  0.11&  5.92$\pm$  0.07&\\
     411& 28.01$\pm$  0.17&  7.44$\pm$  0.05&     678& 28.48$\pm$  0.19&  5.37$\pm$  0.03&     630$^*$& 27.31$\pm$  0.10&  5.44$\pm$  0.04&    1268& 28.37$\pm$  0.17&  5.20$\pm$  0.08&\\
     422& 28.35$\pm$  0.24&  5.55$\pm$  0.03&     726& 26.76$\pm$  0.06&  8.32$\pm$  0.07&     636& 28.19$\pm$  0.23&  5.34$\pm$  0.18&    1388& 27.58$\pm$  0.17&  7.29$\pm$  0.25&\\
     425& 26.76$\pm$  0.07&  5.24$\pm$  0.07&     915& 27.56$\pm$  0.23&  5.31$\pm$  1.93&     640& 26.87$\pm$  0.09&  6.02$\pm$  0.02&    1428& 28.34$\pm$  0.17&  6.89$\pm$  0.07&\\
     437& 28.67$\pm$  0.22&  5.87$\pm$  0.09&    1024& 28.06$\pm$  0.16&  7.49$\pm$  0.07&     653& 24.65$\pm$  0.01&  5.42$\pm$  0.07&    1434& 28.40$\pm$  0.21&  6.39$\pm$  0.15&\\
     446& 27.60$\pm$  0.13&  6.02$\pm$  0.02&    1074& 26.68$\pm$  0.10&  5.78$\pm$  2.23&     774& 28.07$\pm$  0.16&  5.93$\pm$  0.09&    1494& 26.11$\pm$  0.05&  5.76$\pm$  0.09&\\
     466& 27.12$\pm$  0.14&  5.75$\pm$  0.02&    1105& 27.90$\pm$  0.17&  5.29$\pm$  2.01&     790& 25.52$\pm$  0.02&  5.78$\pm$  0.11&    1513& 28.75$\pm$  0.32&  8.54$\pm$  0.01&\\
     475& 27.82$\pm$  0.19&  5.10$\pm$  0.10&    1164& 28.55$\pm$  0.24&  5.96$\pm$  0.09&     797& 25.31$\pm$  0.04&  5.97$\pm$  2.29&    1529& 27.38$\pm$  0.10&  5.08$\pm$  0.06&\\
     491& 28.58$\pm$  0.25&  5.16$\pm$  0.05&    1260& 28.59$\pm$  0.21&  5.24$\pm$  0.01&     813& 28.20$\pm$  0.18&  5.90$\pm$  0.09&    1733& 26.82$\pm$  0.20&  8.76$\pm$  0.85&\\
     535& 28.09$\pm$  0.25&  5.12$\pm$  0.49&    1333& 26.68$\pm$  0.06&  5.16$\pm$  0.09&     880& 27.04$\pm$  0.12&  5.19$\pm$  0.59&    1751& 28.12$\pm$  0.41&  8.32$\pm$  0.04&\\
     548$^*$& 28.42$\pm$  0.20&  8.56$\pm$  0.02&    1405& 26.33$\pm$  0.06&  5.16$\pm$  0.04&     922& 26.86$\pm$  0.10&  5.49$\pm$  0.03&    1758& 26.65$\pm$  0.18&  8.96$\pm$  4.24&\\
     560& 29.03$\pm$  0.22&  5.18$\pm$  0.11&    1457& 28.64$\pm$  0.28&  6.07$\pm$  0.24&     955& 24.52$\pm$  0.03&  5.54$\pm$  0.26&    1970& 28.08$\pm$  0.30&  5.41$\pm$  2.38&\\
     561& 26.78$\pm$  0.10&  6.37$\pm$  0.02&    1494& 27.57$\pm$  0.21&  7.08$\pm$  0.03&    1028& 27.67$\pm$  0.12&  5.65$\pm$  0.08&    2014& 27.80$\pm$  0.24&  6.61$\pm$  0.05&\\
     626& 27.48$\pm$  0.09&  5.55$\pm$  0.01&    1589& 27.12$\pm$  0.16&  7.50$\pm$  0.12&    1095& 26.52$\pm$  0.11&  5.73$\pm$  0.07&    2316& 28.81$\pm$  0.27&  7.93$\pm$  0.04&\\
     657& 28.55$\pm$  0.29&  9.33$\pm$  0.07&    1608& 27.24$\pm$  0.23&  5.03$\pm$  0.21&    1178& 26.82$\pm$  0.06&  6.00$\pm$  0.05&    2364& 28.81$\pm$  0.23&  5.71$\pm$  0.06&\\
     707& 29.01$\pm$  0.23&  6.59$\pm$  0.04&    1614& 27.15$\pm$  0.18&  6.29$\pm$  0.21&    1286& 27.41$\pm$  0.19&  5.05$\pm$  0.26&    2368& 26.60$\pm$  0.07&  5.94$\pm$  0.06&\\
     709& 28.27$\pm$  0.27&  6.31$\pm$  0.02&    1632& 28.17$\pm$  0.36&  6.08$\pm$  2.29&    1333& 28.15$\pm$  0.27&  5.22$\pm$  0.33&    2410& 27.10$\pm$  0.10&  6.00$\pm$  0.03&\\
     742& 27.24$\pm$  0.08&  6.55$\pm$  0.26&    1635& 27.24$\pm$  0.20&  5.61$\pm$  2.31&    1363$^*$& 26.50$\pm$  0.09&  5.16$\pm$  0.74&    2535& 26.91$\pm$  0.07&  5.54$\pm$  0.09&\\
     808& 26.60$\pm$  0.07&  5.36$\pm$  0.01&    1660& 26.83$\pm$  0.12&  5.51$\pm$  0.44&    1398& 26.48$\pm$  0.11&  5.17$\pm$  0.07&    2619& 27.09$\pm$  0.07&  5.79$\pm$  0.11&\\
     809& 27.55$\pm$  0.11&  5.42$\pm$  0.04&    1706& 26.91$\pm$  0.11&  5.42$\pm$  0.12&    1481& 26.92$\pm$  0.10&  5.15$\pm$  2.19&    2747& 28.15$\pm$  0.23&  6.17$\pm$  0.04&\\
     834& 26.40$\pm$  0.10&  5.58$\pm$  0.07&    1815& 27.19$\pm$  0.16&  7.22$\pm$  2.78&    1563& 28.31$\pm$  0.20&  6.86$\pm$  0.03&    2764& 27.18$\pm$  0.10&  5.67$\pm$  0.07&\\
     835& 27.69$\pm$  0.18&  6.15$\pm$  0.09&    1827& 27.31$\pm$  0.14&  5.91$\pm$  0.02&    1584& 28.28$\pm$  0.19&  6.22$\pm$  0.04&    2792& 28.14$\pm$  0.16&  7.18$\pm$  0.11&\\
     855& 27.60$\pm$  0.13&  6.02$\pm$  0.03&    1829& 28.65$\pm$  0.20&  5.96$\pm$  0.01&    1622& 27.60$\pm$  0.12&  6.25$\pm$  0.03&    2833& 27.70$\pm$  0.12&  7.26$\pm$  0.09&\\
     863& 27.01$\pm$  0.07&  5.87$\pm$  0.06&    1900& 29.14$\pm$  0.91&  5.17$\pm$  0.30&    1737& 26.89$\pm$  0.13&  6.04$\pm$  0.07&    2950$^*$& 28.19$\pm$  0.21&  8.62$\pm$  3.19&\\
     902& 29.02$\pm$  0.53&  5.21$\pm$  0.06&    1909& 27.86$\pm$  0.21&  5.43$\pm$  2.28&    1772& 24.55$\pm$  0.03&  5.14$\pm$  0.10&    2966& 27.51$\pm$  0.11&  6.40$\pm$  0.03&\\
     921& 28.74$\pm$  0.49&  5.28$\pm$  0.09&    1956& 28.16$\pm$  0.16&  7.81$\pm$  0.04&    1802& 25.87$\pm$  0.05&  5.53$\pm$  0.04&    3027& 28.03$\pm$  0.16&  5.83$\pm$  0.01&\\
     943& 28.18$\pm$  0.22&  6.97$\pm$  0.07&    1997& 27.56$\pm$  0.17&  8.10$\pm$  0.05&    1841& 27.81$\pm$  0.13&  5.74$\pm$  0.05&    3073& 28.13$\pm$  0.18&  5.10$\pm$  0.10&\\
     945& 28.61$\pm$  0.29&  5.62$\pm$  0.19&    2018$^*$& 28.26$\pm$  0.15&  5.31$\pm$  0.84&    1868& 25.49$\pm$  0.02&  5.64$\pm$  2.39&    3162& 27.67$\pm$  0.14&  6.07$\pm$  0.15&\\
    1012& 28.31$\pm$  0.13&  5.24$\pm$  0.41&    2067& 28.19$\pm$  0.23&  5.07$\pm$  0.20&    1874$^*$& 27.27$\pm$  0.11&  5.48$\pm$  0.08&    3195$^*$& 28.71$\pm$  0.48&  6.14$\pm$  3.67&\\
    1020& 29.06$\pm$  0.62&  5.70$\pm$  0.07&    2157& 28.34$\pm$  0.17&  5.36$\pm$  0.09&    2156& 25.26$\pm$  0.05&  5.50$\pm$  2.31&    3236$^*$& 27.12$\pm$  0.23&  9.11$\pm$  1.04&\\
    1028$^*$& 27.62$\pm$  0.20&  7.08$\pm$  0.14&    2169& 28.09$\pm$  0.15&  5.97$\pm$  0.03&    2191& 27.31$\pm$  0.13&  5.46$\pm$  0.26&    3374& 26.95$\pm$  0.12&  7.43$\pm$  0.09&\\
    1032& 28.21$\pm$  0.14&  7.09$\pm$  0.08&    2179& 26.69$\pm$  0.07&  6.25$\pm$  0.03&    2204& 27.19$\pm$  0.07&  5.38$\pm$  2.16&   &        &        &\\
    1051& 27.11$\pm$  0.21&  6.56$\pm$  0.07&    2190& 28.25$\pm$  0.18&  5.40$\pm$  2.19&    2302& 27.01$\pm$  0.14&  5.37$\pm$  0.10&   &        &        &\\
    1273& 27.31$\pm$  0.11&  6.61$\pm$  0.02&    2196& 28.56$\pm$  0.29&  5.91$\pm$  0.06&    2312& 26.45$\pm$  0.05&  6.12$\pm$  0.04&   &        &        &\\
    1333& 27.23$\pm$  0.09&  5.64$\pm$  0.03&    2204& 26.99$\pm$  0.13&  6.30$\pm$  0.03&    2321& 27.76$\pm$  0.23&  5.10$\pm$  0.14&   &        &        &\\
    1387$^*$& 27.29$\pm$  0.11&  6.72$\pm$  0.23&    2236& 27.67$\pm$  0.13&  5.97$\pm$  0.03&    2368$^*$& 26.54$\pm$  0.16&  5.75$\pm$  0.08&   &        &        &\\
    1399& 26.94$\pm$  0.08&  5.11$\pm$  0.01&    2240& 27.96$\pm$  0.07&  5.75$\pm$  0.07&    2429& 28.15$\pm$  0.21&  5.05$\pm$  0.10&   &        &        &\\
    1450& 28.30$\pm$  0.26&  5.48$\pm$  0.25&    2315$^*$& 27.46$\pm$  0.15&  5.13$\pm$  0.06&    2442& 26.66$\pm$  0.08&  5.21$\pm$  0.11&   &        &        &\\
    1516& 28.15$\pm$  0.15&  5.14$\pm$  0.06&    2323& 28.40$\pm$  0.18&  5.18$\pm$  0.03&    2520$^*$& 27.20$\pm$  0.17&  9.14$\pm$  1.04&   &        &        &\\
    1622& 28.94$\pm$  0.39&  5.87$\pm$  0.06&    2324& 28.15$\pm$  0.25&  6.29$\pm$  0.02&    2575& 28.24$\pm$  0.26&  6.17$\pm$  0.23&   &        &        &\\
    1686& 28.06$\pm$  0.16&  5.02$\pm$  0.06&    2337& 27.30$\pm$  0.16&  6.16$\pm$  0.03&    2584& 28.61$\pm$  0.30&  5.17$\pm$  1.83&   &        &        &\\
    1718& 24.03$\pm$  0.01&  6.21$\pm$  0.09&    2385& 28.09$\pm$  0.16&  8.83$\pm$  0.03&    2585& 27.63$\pm$  0.21&  5.69$\pm$  0.12&   &        &        &\\
    1747$^*$& 27.69$\pm$  0.10&  5.20$\pm$  2.53&    2411& 28.65$\pm$  0.20&  5.91$\pm$  0.05&    2625& 27.41$\pm$  0.21&  5.70$\pm$  0.15&   &        &        &\\
    1762& 28.65$\pm$  0.18&  5.25$\pm$  0.01&    2462& 28.39$\pm$  0.18&  5.69$\pm$  0.03&    2656& 28.90$\pm$  0.61&  5.59$\pm$  2.50&   &        &        &\\
    1968$^*$& 27.38$\pm$  0.09&  5.19$\pm$  0.08&    2554& 27.81$\pm$  0.21&  5.90$\pm$  0.03&    2667$^*$& 26.83$\pm$  0.09&  5.27$\pm$  0.37&   &        &        &\\
    1990& 27.99$\pm$  0.21&  7.07$\pm$  0.13&    2555& 27.52$\pm$  0.20&  6.00$\pm$  0.02&    2730& 26.90$\pm$  0.12&  5.36$\pm$  0.19&   &        &        &\\
    2002& 28.25$\pm$  0.16&  6.46$\pm$  0.02&   &        &        &    2745& 26.83$\pm$  0.12&  5.06$\pm$  0.24&   &        &        &\\
    2007& 28.70$\pm$  0.29&  5.82$\pm$  0.07&   &        &        &    2782& 27.38$\pm$  0.23&  5.27$\pm$  2.23&   &        &        &\\
    2036& 26.95$\pm$  0.07&  8.32$\pm$  0.03&   &        &        &    2799& 28.40$\pm$  0.26&  5.22$\pm$  0.03&   &        &        &\\
    2037& 28.22$\pm$  0.15&  5.08$\pm$  0.07&   &        &        &    2840& 28.04$\pm$  0.21&  5.18$\pm$  0.20&   &        &        &\\
    2066$^*$& 27.82$\pm$  0.16&  5.95$\pm$  2.34&   &        &        &    2843& 26.01$\pm$  0.06&  5.47$\pm$  2.23&   &        &        &\\
    2112& 28.45$\pm$  0.17&  5.07$\pm$  0.07&   &        &        &    2852& 27.67$\pm$  0.14&  5.08$\pm$  0.22&   &        &        &\\
    2181& 28.01$\pm$  0.21&  5.17$\pm$  0.14&   &        &        &    2860& 28.02$\pm$  0.25&  5.37$\pm$  0.06&   &        &        &\\
    2202& 27.69$\pm$  0.11&  5.86$\pm$  0.01&   &        &        &    2883& 25.91$\pm$  0.06&  6.37$\pm$  0.04&   &        &        &\\
    2241& 28.33$\pm$  0.16&  6.84$\pm$  0.04&   &        &        &    2902& 27.71$\pm$  0.23&  6.25$\pm$  0.04&   &        &        &\\
    2257& 28.62$\pm$  0.18&  7.53$\pm$  0.43&   &        &        &    3015& 28.53$\pm$  0.25&  5.76$\pm$  0.42&   &        &        &\\
    2261& 27.29$\pm$  0.10&  7.97$\pm$  0.10&   &        &        &    3017& 27.66$\pm$  0.18&  5.04$\pm$  0.69&   &        &        &\\
    2287& 27.97$\pm$  0.16&  8.50$\pm$  0.94&   &        &        &    3066& 28.62$\pm$  0.25&  6.24$\pm$  0.01&   &        &        &\\
    2316& 27.98$\pm$  0.19&  7.66$\pm$  0.02&   &        &        &    3067& 27.21$\pm$  0.15&  5.97$\pm$  0.12&   &        &        &\\
    2325& 28.54$\pm$  0.18&  5.36$\pm$  0.03&   &        &        &    3076$^*$& 27.36$\pm$  0.14&  9.19$\pm$  1.04&   &        &        &\\
    2338& 28.86$\pm$  0.22&  6.87$\pm$  0.04&   &        &        &   &        &        &   &        &        &\\
    2346& 26.78$\pm$  0.06&  7.79$\pm$  0.04&   &        &        &   &        &        &   &        &        &\\
    2380& 27.71$\pm$  0.22&  7.93$\pm$  0.14&   &        &        &   &        &        &   &        &        &\\
    2388& 27.57$\pm$  0.21&  5.17$\pm$  0.09&   &        &        &   &        &        &   &        &        &\\
    2434& 28.40$\pm$  0.16&  5.82$\pm$  0.07&   &        &        &   &        &        &   &        &        &\\
    2446& 28.05$\pm$  0.15&  5.73$\pm$  0.08&   &        &        &   &        &        &   &        &        &\\
    2452& 27.00$\pm$  0.09&  5.74$\pm$  0.07&   &        &        &   &        &        &   &        &        &\\
    2471& 28.72$\pm$  0.18&  5.66$\pm$  0.07&   &        &        &   &        &        &   &        &        &\\
    2544& 27.03$\pm$  0.11&  5.26$\pm$  0.08&   &        &        &   &        &        &   &        &        &\\
    2567$^*$& 28.98$\pm$  0.27&  5.10$\pm$  0.15&   &        &        &   &        &        &   &        &        &\\
    2595& 27.14$\pm$  0.10&  6.33$\pm$  0.05&   &        &        &   &        &        &   &        &        &\\
\bottomrule 
\label{tb:gal_list}
\end{longtable}
\begin{samepage}
\begin{tablenotes}
\item[*]{Some objects with possibly problematic SEDs are marked with ``*". They are mostly objects showing some flux below the Lyman break, plus some sources detected only in one band. However, we verified that there are no solid reasons to remove them and the photometric redshift solutions appear to be reliable. In particular, the cases of detected flux below the break are mostly due to some contamination from nearby sources or to noise or background fluctuations. }
\end{tablenotes}
\end{samepage}
\end{ThreePartTable} 

\clearpage

\section{The influence of photometric errors}\label{photo_errors}

\begin{figure}
\centering{
\includegraphics[width=0.5\textwidth]{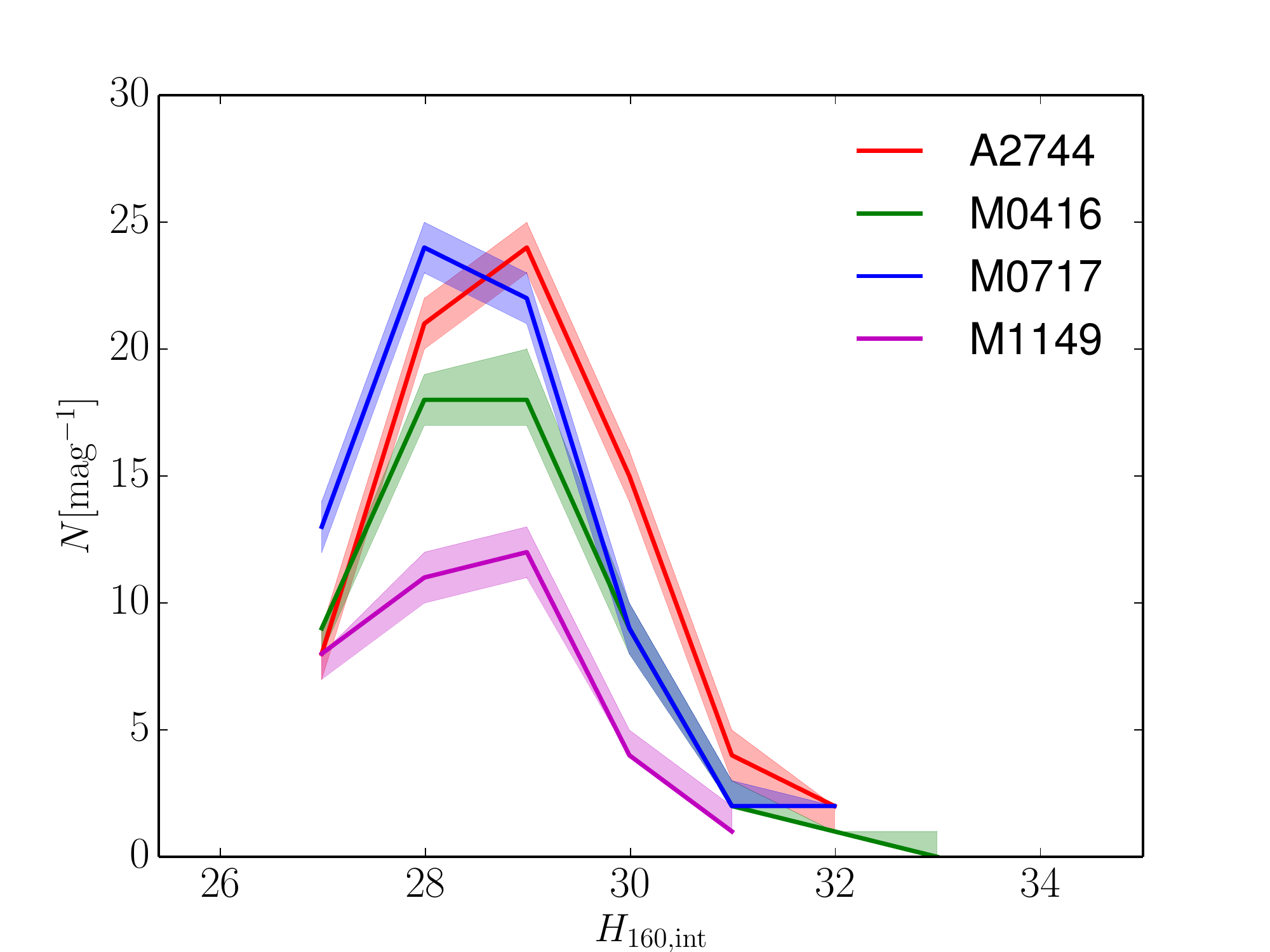}
\caption {The number counts and $1\sigma$ variance induced by the photometric errors for the four FFs clusters.}
\label{fig:photo_errors}
}
\end{figure}

In this section we investigate the influence of the photometric errors on the final constraints on the turn-over magnitude.
For each galaxy with $5.0 < z <7.0$, according to its $H_{160}$ and photometric error, we randomly assign a new $H_{160}$ from the Gaussian probability distribution.
We then get the corresponding new demagnified magnitude and build new number counts. For each galaxy we make 10000 random realizations and finally we have the 10000 number count realizations.  Based on these realizations we get the corresponding $1\sigma$ variance of the number counts rise from the photometric errors.  The results for the four FFs clusters are shown in Fig. \ref{fig:photo_errors}. 

We then obtain the new constraints on the turn-over magnitude by using the $1\sigma$ lower and upper limit of the number counts respectively. Using the lower limit, we have $M_{\rm UV}^{\rm T} > -14.8$ and $M_{\rm UV}^{\rm T}>-15.4)$  at 1 and 2 $\sigma$ C.L. respectively; using the upper limit, we have $M_{\rm UV}^{\rm T} > -14.1$ and $M_{\rm UV}^{\rm T} > -14.8$ respectively at 1 and 2 $\sigma$ C.L.. We conclude that the influence of the photometric errors on the final constraints is modest.

\end{appendix}
  
\end{document}